\documentclass[10pt, final, twocolumn,twoside]{IEEEtran}
\usepackage{cite,acronym,lineno}
\usepackage{amssymb,psfrag,graphicx,subfigure,textcomp,bm}
\usepackage{dsfont,stfloats}
\usepackage[dvips]{color}

\usepackage[cmex10]{amsmath}
\newcounter{MYtempeqncnt}

\newcommand\Matrix[2]
{   \left[ \begin{array}{#1}
        #2
    \end{array} \right]
}

\newtheorem{defi}{Definition}
\newtheorem{thm}{Theorem}
\newtheorem{cor}{Corollary}
\newtheorem{lem}{Lemma}
\newtheorem{pro}{Proposition}
\newtheorem{rem}{Remark}

\newcommand\A[2]  { \alpha_{#1}^{(#2)} }
\newcommand\D[2]  { \tau_{#1}^{(#2)}  }
\newcommand\BB[2]  { b_{#1}^{(#2)}  }

\newcommand\gp  {f}
\newcommand\gkj {f}
\newcommand\gk  {f}

\newcommand\E   {  \mathbb{E} }
\newcommand\T   {  \textrm{tr} }
\newcommand\PEB {  \mathcal{P} }
\newcommand\NA  { { \mathcal{N}_\Tx{a} } }
\newcommand\Na  { { N_\Tx{a} }}
\newcommand\NB { { \mathcal{N}_\Tx{b} } }
\newcommand\Nb { { {N}_\Tx{b} } }

\newcommand\V[1]    {  \mathbf{#1} }
\newcommand\B[1]    {  \boldsymbol{#1} }
\newcommand\JTH {  \V{J}_{\B\theta} }

\newcommand\Tx[1]   {   \textrm{#1} }
\newcommand\Th[1]   {   {#1}\textrm{th} }

\newcommand\Diag[1]
{   \Tx{diag} \left\{ #1 \right\}}

\newcommand\JE[1]   {  {\V{J}_\Tx{e}({#1})} }

\newcommand\JA[1]   {  {\V{J}^\Tx{A}_\Tx{e}(#1)} }
\newcommand\bJA[1]   {  {\bar{\V{J}}^\Tx{A}_\Tx{e}(#1)} }
\newcommand\JC[1]   {  {\V{J}^\Tx{C}_\Tx{e}(#1)} }

\newcommand\JU[1]   {  \V{J}^\Tx{U}_\Tx{e}(#1) }
\newcommand\JL[1]   {  \V{J}^\Tx{L}_\Tx{e}(#1) }

\newcommand\C   {  \V{C} }

\newcommand\U   {  \V{U} }
\newcommand\q   {  \V{q} }
\newcommand\R   {  \V{J}_\Tx{r} }

\newcommand\GP  { \B\Xi_{\V{P}} }

\newcommand\LrT
{ f(\V{r},\B\theta)}

\begin{document}



\title{Fundamental Limits of Wideband Localization---\\Part II:
Cooperative Networks}

\author{Yuan~Shen,~\IEEEmembership{Student~Member,~IEEE,} Henk~Wymeersch,~\IEEEmembership{Member,~IEEE,} and Moe~Z.~Win,~\IEEEmembership{Fellow,~IEEE}
\thanks{Manuscript received June 20, 2009; revised February 18, 2010. Current version published Month Day 2010. This research was supported, in part, by the National Science Foundation under Grant ECCS-0901034, the Office of Naval Research Presidential Early Career Award for Scientists and Engineers (PECASE) N00014-09-1-0435, and MIT Institute for Soldier Nanotechnologies. The  paper was presented in part at the IEEE Wireless Communications and Networking Conference, Hong Kong, March, 2007, and the IEEE International Symposium on Spread Spectrum Techniques \& Applications, Bologna, Italy, August, 2008.} 
\thanks{Y.~Shen and M.~Z.~Win are with the Laboratory for Information and Decision Systems (LIDS), Massachusetts Institute of Technology, 77 Massachusetts Avenue, Cambridge, MA 02139 USA (e-mail: \{shenyuan, moewin\}@mit.edu).}
\thanks{H.~Wymeersch was with the Laboratory for Information and Decision Systems (LIDS), Massachusetts Institute of Technology, and is now with the Department of Signals and Systems, Chalmers University of Technology, Gothenburg, Sweden. (e-mail: {henkw@chalmers.se}).}
\thanks{Communicated by Massimo Franceschetti, Associate Editor for Communication Networks.}
\thanks{Color versions of the figures in this paper are available online at http://ieeexplore.ieee.org.}
\thanks{Digital Object Identifier XXX.XXX}
}


\maketitle

\markboth{IEEE Transactions on Information Theory, Vol.~X, No.~Y, Month~2010}{Shen \emph{\MakeLowercase{et al.}}: Fundamental Limits of Wideband Localization---Part II: Cooperative Networks}

\begin{abstract}
The availability of positional information is of great importance in many commercial, governmental, and military applications. Localization is commonly accomplished through the use of radio communication between mobile devices (agents) and fixed infrastructure (anchors). However, precise determination of agent positions is a challenging task, especially in harsh environments due to radio blockage or limited anchor deployment. In these situations, cooperation among agents can significantly improve localization accuracy and reduce localization outage probabilities. A general framework of analyzing the fundamental limits of wideband localization has been developed in Part I of the paper. Here, we build on this framework and establish the fundamental limits of wideband cooperative location-aware networks. Our analysis is based on the waveforms received at the nodes, in conjunction with Fisher information inequality. We provide a geometrical interpretation of equivalent Fisher information for cooperative networks.
This approach allows us to succinctly derive fundamental performance limits and their scaling behaviors, and to treat anchors and agents in a unified way from the perspective of localization accuracy. Our results yield important insights into how and when cooperation is beneficial.
\end{abstract}


%
%

\section{Introduction}\label{sec:Intr}

\begin{figure}[t]
    \begin{center}
		    \psfrag{1}[c][][1.3]{1}
		\psfrag{2}[c][][1.3]{2}
		    \psfrag{C}[c][][1.3]{C}
		    \psfrag{D}[c][][1.3]{ D}
	    \psfrag{A}[c][][1.3]{A}
	    \psfrag{B}[c][][1.3]{B}
\includegraphics[angle=0,width=0.75\linewidth,draft=false]{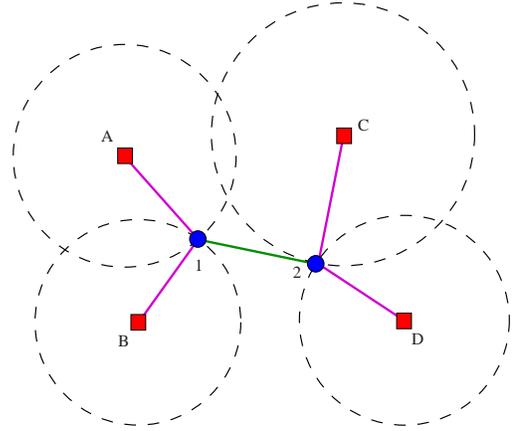}
    \caption{\label{fig:Sen_Net} Cooperative localization: the anchors (A, B, C, and D) communicate with the agents (1 and 2).  agent 1 is not in the communication/ranging range of anchor C and D, while agent 2 is not in the communication/ranging range of anchor A and B. Neither agents can trilaterate its position based solely on the information from its neighboring anchors. However, cooperation between agent 1 and 2 enables both agents to be localized.
    }
    \end{center}
\end{figure}

The availability of absolute or relative positional information is
of great importance in many applications, such as localization
services in cellular networks, search-and-rescue operations, asset
tracking, blue force tracking, vehicle routing, and intruder
detection \cite{SayTarKha:05, PahLiMak:02, CafStu:98, ChoKum:03, PatAshKypHerMosCor:05, WymLieWin:J09, GezTiaGiaKobMolPooSah:05, JouDarWin:J08}. Location-aware networks generally consist of two kinds of nodes: anchors and agents (see Fig.~\ref{fig:Sen_Net}), where anchors have known positions while agents have unknown positions. Conventionally, each agent localizes itself based on range measurements from at least three distinct anchors (in two-dimensional localization). Two common examples include the global positioning system (GPS) \cite{Kap:96,Spi:78} and beacon localization \cite{ZagParBusMel:98,HigLamSmi:06}. In GPS, an agent can determine its location based on the signals received from a constellation of GPS satellites. However, GPS does not operate well in harsh environments, such as indoors or in urban canyons, since the signals cannot propagate through obstacles \cite{Kap:96,JouDarWin:J08, GezTiaGiaKobMolPooSah:05}. Beacon localization,
on the other hand, relies on terrestrial anchors, such as WiFi
access points or GSM base stations  \cite{ZagParBusMel:98,HigLamSmi:06}. However, in areas where network
coverage is sparse, e.g., in emergency situations, localization
errors can be unacceptably large.

Conventionally, high-accuracy localization can only be achieved
using high-power anchors or a high-density anchor deployment, both
of which are cost-prohibitive and impractical in realistic settings.
Hence, there is a need for localization systems that can
achieve high accuracy in harsh environments with limited
infrastructure requirements \cite{WymLieWin:J09, JouDarWin:J08, GezTiaGiaKobMolPooSah:05}. A practical way to address this need is through a combination of \emph{wideband transmission} and \emph{cooperative localization}. 
The fine delay resolution and robustness of wide bandwidth or ultra-wide bandwidth (UWB) transmission enable accurate and reliable 
range (distance) measurements in harsh environments  \cite{WinSch:J00, CasWinMol:J02, LeeSch:02, DarConFerGioWin:J09, ZhaLawGua:05, LowCheLawNgLee:05}.\footnote{Other aspects of UWB technology can be found in \cite{CasWinVatMol:J07, QueWinDar:J07, YanGia:04a, RidWin:J06, SuwWin:J07, Mol:05, Mol:09}.} Hence, these transmission techniques are particularly well-suited for localization.
Cooperative localization is an emerging paradigm that circumvents the needs for high-power, high-density anchor deployment, and offers additional localization accuracy by enabling the agents to help each other in estimating their positions \cite{PatAshKypHerMosCor:05, WymLieWin:J09, SavRabBeu:01, ChaSah:04, Lar:04}. In Fig.~\ref{fig:Sen_Net}, for example, since each agent is in the communication/ranging range of only two anchors, neither agents can trilaterate its position based solely on the information from its neighboring anchors. However, cooperation enables both agents to be localized.



Understanding the fundamental limits of localization is crucial not
only for providing a performance benchmark but also for guiding the
deployment and operation of location-aware networks. Localization
accuracy is fundamentally limited due to random phenomena such as
noise, fading, shadowing, and multipath propagation. The impact of
these phenomena has been investigated for non-cooperative
localization \cite{GezTiaGiaKobMolPooSah:05, JouDarWin:J08,
SheWin:J10a, SheWin:J10, QiKobSud:06}. However, little is known
regarding the bounds for cooperative localization. In particular,
bounds on the cooperative localization performance were previously
derived in \cite{ChaSah:04, Lar:04} using only specific ranging
models. In other words, these works start from signal metrics,
extracted from the received waveforms.\footnote{Commonly used signal
metrics include time-of-arrival (TOA) \cite{JouDarWin:J08, LeeSch:02,
Mai:08, GezTiaGiaKobMolPooSah:05, ZhaLawGua:05},
time-difference-of-arrival (TDOA) \cite{Caf:00, RapReeWoe:96},
angle-of-arrival (AOA) \cite{NicNat:03,GezTiaGiaKobMolPooSah:05},
and received signal strength (RSS)
\cite{PatHer:03,GezTiaGiaKobMolPooSah:05, PavCosMazConDar:06}.} Such a process may discard information relevant for localization.
Furthermore, the statistical models for those signal metrics depend
heavily on the measurement processes. For instance, the ranging
error of the time-of-arrival (TOA) metric is commonly modeled as
additive Gaussian \cite{ChaSah:04, Lar:04, QiKobSud:06}. However, other studies (both theoretical
\cite{LeeSch:02, HamSch:92, LeeYoo:05} and experimental \cite{JouDarWin:J08, LowCheLawNgLee:05})
indicate that the ranging error is not Gaussian. Hence, when
deriving the fundamental limits of localization accuracy, it is
important to start from the received waveforms rather than from
signal metrics extracted from those waveforms.

In Part I \cite{SheWin:J10a}, we have developed a general framework to characterize the localization accuracy of a given agent. In this paper, we build on the framework and determine fundamental properties of \emph{cooperative} location-aware networks employing wideband transmission. The main contributions of this paper are as follows:

\begin{itemize}
    \item
        We derive the fundamental limits of localization accuracy
        for wideband wireless {cooperative} networks in terms of a
        performance measure called the \emph{squared position error bound} (SPEB).
    \item
        We employ the notion of \emph{equivalent Fisher information} (EFI) to derive the network localization information, and
        show that this information
        can be decomposed into basic building blocks associated with
        every pair of the nodes, called the \emph{ranging information} (RI). 
    \item
        We quantify the contribution of the {a priori} knowledge
        of the channel parameters and the agents' positions to
        the network localization information, and show that agents
        and anchors can be treated in a \emph{unified way}: anchors are special agents with infinite {a priori} position
        knowledge.
    \item
        We put forth a \emph{geometric interpretation} of the EFI matrix (EFIM) using eigen-decomposition, providing insights into the network localization problem.
    \item
        We derive \emph{scaling laws} for the SPEB for both dense and extended location-aware networks, characterizing the behavior of cooperative location-aware
        networks in an asymptotic regime.
\end{itemize}
The proposed framework generalizes the existing work on non-cooperative localization \cite{SheWin:J10a} to cooperative networks, provide insights into the network localization problem, and can guide the design and deployment of location-aware networks.


The rest of the paper is organized as follows. Section
\ref{Sec:Model} presents the system model and the concept of SPEB. In
Section \ref{Sec:Eval}, we apply the notion of EFI to derive the SPEB. Then in Section \ref{Sec:Geom}, we provide a
geometric interpretation of EFIM for localization and derive
scaling laws for the SPEB. Finally, numerical results are given in
Section \ref{sec:Simu}, and conclusions are drawn in the last
section. 

\subsubsection*{Notation}
The notation $\E_{\V{x}}\{\cdot\}$ is the expectation operator with
respect to the random vectors $\V{x}$; $\V{A} \succ \V{B}$ and $\V{A} \succeq \V{B}$ denote that the matrix $ \V{A} - \V{B} $ is positive definite and positive semi-definite, respectively; $\T\{\cdot\}$ denotes the trace of a square matrix; $\left[\, \cdot \, \right]^\text{T}$ denotes the transpose of its argument; $\left[\, \cdot \, \right]_{n \times n,k}$ denotes the $\Th{k}$ $n \times n$ submatrix that starts from element $n(k-1)+1$ on the diagonal of its argument; $\left[\, \cdot \, \right]_{r_1:r_2,c_1:c_2}$ denotes a submatrix composed of the rows $r_1$ to $r_2$ and the columns $c_1$ to $c_2$ of its argument; and $\|\cdot\|$ denotes the Euclidean norm of its argument. We also denote by $f(\V{x})$ the probability density function (PDF) $f_\V{X}(\V{x})$ of the random vector $\V{X}$ unless specified otherwise.

%
%

\section{System Model}\label{Sec:Model}

In this section, we describe the wideband channel model and formulate the localization problem. We briefly review the information inequality and the performance measure called SPEB.

\subsection{Signal Model}\label{Sec:Model_SysMod}

Consider a synchronous network consisting of $\Nb$ anchors (or beacons) and $\Na$ agents with fixed topology.\footnote{We consider synchronous networks for notional convenience. Our approach is also valid for asynchronous networks, where devices employ round-trip time-of-flight measurements \cite{Mol:B05, Mol:09}.} Anchors have perfect knowledge of their positions, while each agent attempts to estimate its position based on the waveforms received from neighboring nodes (see Fig.~\ref{fig:Sen_Net}). Unlike conventional localization techniques, we consider a cooperative setting, where agents utilize waveforms received from neighboring agents in addition to those from anchors. The set of agents is denoted by $\NA = \left\{ 1,2, \dotsc, \Na \right\}$, while the set of anchors is $\NB= \left\{\Na+1,\Na+2, \dotsc, \Na+\Nb \right\}$. The position of node $k$ is denoted by $\V{p}_k \triangleq [\,x_k \;\;  y_k\,]^\text{T}$.\footnote{For convenience, we focus on two-dimensional localization where $\V{p}_k \in \mathbb{R}^2$, and we will later mention extensions to three-dimensional localization.} Let $\phi_{kj}$ denote the angle from node $k$ to node $j$, i.e.,
    \begin{align*}
        \phi_{kj} = \tan^{-1} \frac{y_k-y_j}{x_k-x_j}\,,
    \end{align*}
and $\V{q}_{kj} \triangleq [\,\cos\phi_{kj} \;\; \sin\phi_{kj}
\,]^\text{T}$ denote the corresponding unit vector. 

The received waveform at the $\Th{k}$ agent ($k \in \NA$) from the
$\Th{j}$ node ($j \in \NB \cup \NA\backslash\{k\}$) can be written
as \cite{Mol:05,SalVal:87}
    \begin{align}\label{eq:Model_signal}
        r_{kj}(t) = \sum_{l=1}^{L_{kj}} \A{kj}{l} \, s\left(t - \D{kj}{l}\right)
            + z_{kj}(t)\, ,  \quad  t \in [ \, 0, T_{\text{ob}})
        \, ,
    \end{align}
where $s(t)$ is a known wideband waveform with Fourier transform
$S(f)$, $\A{kj}{l}$ and $\D{kj}{l}$ are the amplitude and delay,
respectively, of the $\Th{l}$ path,\footnote{We consider the general
case where the wideband channel is not necessarily reciprocal. Our
results can be easily specialized to the reciprocal case, where we
have $L_{kj}=L_{jk}$, $\A{kj}{l} = \A{jk}{l}$, and $\D{kj}{l} =
\D{jk}{l}$ hence $\BB{kj}{l} = \BB{jk}{l}$, for $ l = 1, 2, \ldots,
L_{kj}$.} $L_{kj}$ is the number of multipath
components, 
$z_{kj}(t)$ represents the observation noise, modeled as additive
white Gaussian processes with two-sided power spectral density
$N_0/2$, and $[ \, 0, T_{\text{ob}})$ is the observation interval.
The relationship between the positions of nodes and the delays of the propagation paths is
    \begin{align}\label{eq:Model_Tau_L}
        \D{kj}{l} = \frac{1}{c} {\Big[} \, \left\| \V{p}_k - \V{p}_j \right\| + \BB{kj}{l} \, {\Big]},
        \quad j \in \NB \cup \NA\backslash\{k\}
        \,,
    \end{align}
where $c$ is the propagation speed of the signal, and $\BB{kj}{l}\geq 0$ is a
range bias induced by non-line-of-sight (NLOS) propagation.
Line-of-sight (LOS) signals occur when the direct path between nodes
$k$ and $j$ is unobstructed, such that $\BB{kj}{1} = 0$.


\subsection{Error Bounds on Position Estimation}

We first introduce $\bm\theta$ as the vector of unknown parameters,
    \begin{align*}
        \bm\theta = \Matrix{ccccc} { \V{P}^\text{T} & \tilde{\bm\theta}_1^\text{T}
        & \tilde{\bm\theta}_2^\text{T} & \cdots & \tilde{\bm\theta}_{\Na}^\text{T}}^\text{T} ,
    \end{align*}
where $\V{P}$ consists of all the agents' positions
    \begin{align*}
        \V{P} = \Matrix{ccccc} { \V{p}_1^\text{T}
        & \V{p}_2^\text{T} & \cdots &  \V{p}_\Na^\text{T}}^\text{T} ,
    \end{align*}
and $\tilde{\bm\theta}_k$ is the vector of the multipath parameters
associated with the waveforms received at the $\Th{k}$
agent\footnote{In cases where the channel is reciprocal,
only half of the multipath parameters are needed. Without loss of generality, we only use ${\Big\{}
\A{kj}{l},\BB{kj}{l} :  \; k,j \in \NA, k > j {\Big\}}$.}
    \begin{align*}
        \tilde{\bm\theta}_k
        = \Matrix{ccccccc}{\bm\kappa_{k,1}^\text{T} & \cdots & \bm\kappa_{k,k-1}^\text{T} &
        \bm\kappa_{k,k+1}^\text{T} & \cdots & \bm\kappa_{k,\Na+\Nb}^\text{T}}^\text{T}
        ,
    \end{align*}
in which $\bm\kappa_{kj}$ is the vector of the multipath parameters
associated with $r_{kj}(t)$,\footnote{The bias $\BB{kj}{1}=0$ for
LOS signals. From the perspective of Bayesian estimation, it can
be thought of as a random parameter with infinite {a priori} Fisher
information \cite{SheWin:J10a}.}
    \begin{align*}
        \bm\kappa_{kj} = \Matrix{ccccccc}{\BB{kj}{1} & \A{kj}{1}
        &   \cdots & \BB{kj}{L_{kj}} & \A{kj}{L_{kj}} }^\text{T} .
    \end{align*}

Secondly, we introduce $\V{r}$ as the vector representation of all
the received waveforms, given by $\V{r} =[\,
\V{r}_1^\text{T}\;\; \V{r}_2^\text{T} \;\; \cdots \;\; \V{r}_{\Na}^\text{T}]^\text{T}$, where
    \begin{align*}
        \V{r}_k = \Matrix{cccccc} {\V{r}_{k,1}^\text{T}   &  \cdots & \V{r}_{k,k-1}^\text{T}  & \V{r}_{k,k+1}^\text{T} &  \cdots &  \V{r}_{k, \Na+\Nb}^\text{T}}^\text{T} ,
    \end{align*}
in which $\V{r}_{kj}$ is obtained from the Karhunen-Lo\`{e}ve (KL) expansion of $r_{kj}(t)$ \cite{Tre:68, Poo:B94}. We tacitly assume that when nodes $j$ and $k$ cannot communicate directly, the corresponding entry $\V{r}_{kj}$ is omitted in $\V{r}$.

We can now introduce an estimator $\hat{\B{\theta}}$ of the unknown
parameter $\bm\theta$ based on the observation $\V{r}$. The mean squared error (MSE) matrix of $\hat{\B{\theta}}$ satisfies the
information inequality \cite{Tre:68, ReuMes:97, Poo:B94}
    \begin{align}\label{eq:Model_CRLB_FIM}
        \E_{\V{r},\B\theta} \left\{ {(\hat{\B{\theta}}-{\B\theta})(\hat{\B\theta}-{\B\theta})}^\text{T} \right\} \succeq \JTH^{-1} ,
    \end{align}
where $\JTH$ is the Fisher information matrix (FIM) for $\B\theta$,\footnote{With a slight abuse of notation, $\E_{\V{r},\bm\theta}\{\cdot\}$ in \eqref{eq:Model_CRLB_FIM} and \eqref{eq:Eval_JTH} will be used for deterministic, random, and hybrid cases, with the understanding that the expectation operation is not performed over the deterministic components of $\bm\theta$ \cite{Poo:B94,ReuMes:97}. Note also that for the deterministic components, the lower bound is valid for their unbiased estimates.} given by
    \begin{align}\label{eq:Eval_JTH}
        \JTH & = \E_{\V{r},\bm\theta} \left\{
        - \frac{\partial^2}{\partial \bm\theta \partial \bm\theta^\text{T}}
        \ln f(\V{r},\bm\theta) \right\} ,
    \end{align}
in which $f(\V{r},\bm\theta)$ is the joint PDF of the observation
$\V{r}$ and the parameter vector $\B\Theta$. For an estimate $\hat{\V{p}}_k$ of the $k$th agent's position, equation
\eqref{eq:Model_CRLB_FIM} implies that
    \begin{align*}
        \E_{\V{r}, \bm\theta} \left\{ ( \hat{\V{p}}_k - \V{p}_k )
        (\hat{\V{p}}_k - \V{p}_k )^\text{T} \right\}
        \succeq  \left[ \JTH^{-1} \right]_{2 \times 2,k} .
    \end{align*}

One natural measure for position accuracy is the average squared
position error $\E_{\V{r},\bm\theta} \left\{ \| \hat{\V{p}}_k -
\V{p}_k \|^2 \right\}$, which can be bounded below by
$\PEB(\V{p}_k)$ defined in the following.

\begin{defi}[Squared Position Error Bound \cite{SheWin:J10a}]
The squared position error bound (SPEB) of the $\Th{k}$ agent is defined to be
    \begin{align*}
        \PEB(\V{p}_k) \triangleq \T
        \left\{ \left[ \JTH^{-1} \right]_{2 \times 2, k} \right\} .
    \end{align*}
\end{defi}

Since the error of the position estimate $\hat{\V{p}}_k-\V{p}_k$ is a vector, it may also be of interest to know the position error in a particular direction. The directional position error along a given unit vector $\V{u}$ is the position error projected on it, i.e., $\V{u}^\text{T} (\hat{\V{p}}_k-\V{p}_k)$, and its average squared error $\E_{\V{r},\bm\theta} \left\{
\|\V{u}^\text{T}( \hat{\V{p}}_k - \V{p}_k) \|^2 \right\}$  can be bounded below by $\PEB(\V{p}_k;\V{u})$
defined in the following.\footnote{In higher dimensions, this notion
can be extend to the position error in any subspaces, such as a
hyperplane.}


\begin{defi}[Directional Position Error Bound]
The directional position error bound (DPEB) of the $\Th{k}$ agent with constraint $\V{u}_\bot^\text{T} \, (\hat{\V{p}}_k-\V{p}_k)=0$ is defined to be
    \begin{align*}
        \PEB(\V{p}_k;\V{u}) \triangleq \V{u}^\text{T} \, \left[ \JTH^{-1} \right]_{2 \times 2, k} \, \V{u} \, ,
    \end{align*}
where $\V{u},\V{u}_\bot \in \mathbb{R}^2$ are unit vectors such that
$\langle \V{u}, \V{u}_\bot \rangle = 0$.
\end{defi}

\begin{pro}\label{pro:SPEB_Decop}
The SPEB of the $k$th agent is the sum of the DPEBs in any two orthogonal directions, i.e.,
    \begin{align}\label{eq:Anal_dirPEB}
        \PEB(\V{p}_k) = \PEB(\V{p}_k;\V{u}) + \PEB(\V{p}_k;\V{u}_\bot)
        \, .
    \end{align}
\end{pro}

\begin{IEEEproof}
See Appendix \ref{apd:pro_SPEB_Decop}.
\end{IEEEproof}

\subsection{Joint PDF of Observations and Parameters} \label{Sec:Eval_FIM}

Evaluation of \eqref{eq:Eval_JTH} requires knowledge of the joint
distribution $f(\V{r},\bm\theta)$. We can write $\LrT =
f(\V{r}|{\bm\theta}) \, f(\bm\theta)$, where
$f(\V{r}|{\bm\theta})$ is the likelihood function, and
$f(\bm\theta)$ is the {a priori} distribution of the parameter
$\bm\theta$.\footnote{When a subset of the parameters are
deterministic, they are eliminated from $f(\bm\theta)$.} In this
section, we describe the structure of both functions in detail.

Since the received waveforms $r_{kj}(t)$ are independent conditioned on the parameter $\B\theta$, $f(\V{r}|{\bm\theta})$ can be expressed as \cite{Tre:68, Poo:B94}
    \begin{align}\label{eq:Eval_ConLR_prod}
        f(\V{r}|{\bm\theta}) = \prod_{k \in \NA} \, \prod_{j\in
        \NB\cup\NA\backslash\{k\}} f(\V{r}_{kj}|{\bm\theta}) \, ,
    \end{align}
where
    \begin{align}\label{eq:Eval_ConLR}
        f(\V{r}_{kj}|{\bm\theta}) & \propto \exp {\Bigg\{}
                \frac{2}{N_0} \int_0^{T_\text{ob}} r_{kj}(t)  \sum_{l=1}^{L_{kj}}
                \A{kj}{l}\, s\left(t-\D{kj}{l}\right) dt
                \nonumber \\ 
& \hspace{9mm} - \frac{1}{N_0} \int_0^{T_\text{ob}} \left[\sum_{l=1}^{L_{kj}}
                \A{kj}{l}\, s\left(t-\D{kj}{l}\right)\right]^2 dt {\Bigg\}} .
    \end{align}

When the multipath parameters $\bm\kappa_{kj}$ are independent conditioned on the nodes' positions,\footnote{This is a common model for analyzing wideband communication, unless two nodes are close to each other so that the channels from a third node to them are correlated. Our analysis can also account for the correlated channels, in which case the SPEB will be higher than that corresponding to the independent channels.} $f(\bm\theta)$ can be expressed as
    \begin{align}\label{eq:FIM_Prior_Dis}
        f(\bm\theta) 
		& = \gp(\V{P}) \, \prod_{k \in \NA}
                    \gk(\tilde{\bm\theta}_k | \V{P}) \nonumber \\
        & = \gp(\V{P}) \, \prod_{k \in \NA} \;
                    \prod_{j\in \NB\cup\NA\backslash\{k\}} \gkj(\bm\kappa_{kj} | \V{P})
        \, ,
    \end{align}
where $\gp(\V{P})$ is the joint PDF of all the agents' positions,
and $\gkj(\bm\kappa_{kj} | \V{P})$ is the joint PDF of the multipath
parameters $\bm\kappa_{kj}$ conditioned on the agents' positions.
Based on existing propagation models for wideband and UWB channels
\cite{Mol:09, CasWinMol:J02}, the joint PDF of the channel parameters can be further written as \cite{SheWin:J10a}:
    \begin{align}\label{eq:Anal_PDFkj}
        \gkj(\bm\kappa_{kj} | \V{P}) = \gkj(\bm\kappa_{kj} | d_{kj})
        \, ,
    \end{align}
where $d_{kj} = \|\V{p}_k - \V{p}_j \|$ for $k\in\NA$ and $j\in
\NB\cup\NA\backslash\{k\}$.

Combining \eqref{eq:FIM_Prior_Dis} and
\eqref{eq:Eval_ConLR_prod} leads to
    \begin{align}\label{eq:Eval_LN}
        \ln \LrT & = {\sum_{k \in \NA} \sum_{j\in \NB } {\Big[} \ln
        f(\V{r}_{kj}|{\bm\theta}) + \ln \gkj(\bm\kappa_{kj}|\V{P})} {\Big]} \nonumber \\
        & \hspace{4mm} +  {\sum_{k \in \NA} \sum_{j \in \NA\backslash\{k\}} {\Big[} \ln f(\V{r}_{kj}|{\bm\theta}) + \ln \gkj(\bm\kappa_{kj}|\V{P}) } {\Big]} \nonumber \\
        & \hspace{4mm} +  {\ln \gp(\V{P})}
        \, ,
    \end{align}
where the first and second groups of summation account for the information from anchors and that from agents' cooperation, respectively, and the last term accounts for the information from the a priori knowledge of the agents' positions.
This implies that the FIM for $\bm\theta$ in \eqref{eq:Eval_JTH} can
be written as
$\JTH = \V{J}_{\bm\theta}^\text{A} + \V{J}_{\bm\theta}^\text{C} + \V{J}_{\bm\theta}^\text{P}$, %
%
where $\V{J}_{\bm\theta}^\text{A}$, $\V{J}_{\bm\theta}^\text{C}$, and
$\V{J}_{\bm\theta}^\text{P}$ correspond to the localization information from anchors, agents' cooperation, and {a priori} knowledge of the agents' positions, respectively.

%
%
\begin{figure*}
	[!b] \vspace*{4pt} \hrulefill \normalsize \setcounter{MYtempeqncnt}{\value{equation}} \setcounter{equation}{11} 
    \begin{align}\label{eq:Coop_EFIM}
    \JE{\V{P}} =  \Matrix{cccc}
            {       \JA{\V{p}_1} + \sum\limits_{j \in \NA\backslash\{1\}} \C_{1,j}    &   -\C_{1,2}                             &   \cdots  &   -\C_{1,\Na}  \\
                    -\C_{1,2}            &  \JA{\V{p}_2} + \sum\limits_{j \in \NA\backslash\{2\} } \C_{2,j}  &           &   -\C_{2,\Na}  \\
                    \vdots                  &                              &   \ddots  &      \\
                    -\C_{1,\Na}          &   -\C_{2,\Na}             &      &  \JA{\V{p}_\Na} + \sum\limits_{j \in \NA\backslash\{\Na\}} \C_{\Na,j} }
    \end{align}
	\setcounter{equation}{\value{MYtempeqncnt}} \vspace*{-6pt} 
\end{figure*}

%
%

\section{Evaluation of FIM}\label{Sec:Eval}

In this section, we briefly review the notion of EFI \cite{SheWin:J10a} and apply it to derive the SPEB for each agent. We consider both the cases with and without {a priori} knowledge of the agents' positions. We also introduce the concept of RI, which turns out to be the basic building block for the EFIM.

\subsection{Equivalent Fisher Information Matrix and Ranging Information}

We saw in the previous section that the SPEB can be obtained by
inverting the FIM $\JTH$ in \eqref{eq:Eval_JTH}. However, $\JTH$ is
a matrix of very high dimensions, while only a much smaller submatrix
$\left[ \JTH^{-1} \right]_{2\Na \times 2\Na}$ is of interest. To gain insights into localization problem, we will employ the notions of EFIM and RI \cite{SheWin:J10a}. For the completeness of the paper, we briefly review the notions in the following.

\begin{defi}[Equivalent Fisher Information Matrix]\label{def:EFIM}
Given a parameter vector $\bm\theta = [\, \bm\theta_1^\text{T} \;
\bm\theta_2^\text{T} \,]^\text{T}$ and the FIM $\JTH$ of the form
    \begin{align*}
        \JTH = \Matrix{cc}
                {   \V{A}          &   \V{B}   \\
                    \V{B}^\text{T} &   \V{C}   },
    \end{align*}
where $\bm\theta \in \mathbb{R}^N$, $ \bm\theta_1 \in \mathbb{R}^n$,
$\V{A} \in \mathbb{R}^{n \times n}$, $\V{B} \in \mathbb{R}^{n\times
(N-n)}$, and $\V{C} \in \mathbb{R}^{(N-n) \times (N-n)}$ with $n<N$,
the equivalent Fisher information matrix (EFIM) for $\bm\theta_1$ is
given by
    \begin{align}\label{eq:Sing_EFIM_FIM}
        \JE{\bm\theta_1} \triangleq \V{A} - \V{B} \V{C}^{-1} \V{B}^\text{T}
        \, .
    \end{align}
\end{defi}

Note that the EFIM retains all the necessary information to derive
the information inequality for the parameter $\bm\theta_1$, in a
sense that $[\,\JTH ^{-1}]_{n \times n} = \left[\, \JE{\bm\theta_1}
\, \right]^{-1}$, 
so that the MSE matrix of the estimates of $\bm\theta_1$ is ``bounded'' below by $\left[\,
\JE{\bm\theta_1} \,\right]^{-1}$. The right-hand side of (\ref{eq:Sing_EFIM_FIM}) is known as the Schur's complement of matrix $\V{A}$ \cite{HorJoh:B85}, and it has been used for simplifying the CRBs \cite{QiKobSud:06, Mai:08, BotHosFat:04}.

\begin{defi}[Ranging Information]
The ranging information (RI) is a $2 \times 2$ matrix of the form
$\lambda \, \R(\phi)$, where $\lambda$ is a nonnegative number
called the ranging information intensity (RII) and the matrix
$\R(\phi)$ is called the ranging direction matrix (RDM) with the
following structure:
    \begin{align*}
        \R (\phi) \triangleq \Matrix{cc} {  \cos^2 \phi               & \cos \phi \sin \phi \\
                                            \cos \phi \sin \phi       & \sin^2 \phi    } .
    \end{align*}
\end{defi}

The RDM $\R(\phi)$ has exactly one non-zero eigenvalue equal to $1$
with corresponding eigenvector $\q =\left[\,\cos\phi \;\; \sin\phi
\,\right]^\text{T}$, i.e., $\R(\phi)= \q\, \q^\text{T}$. Thus,
the corresponding RI is ``one-dimensional'' along the direction $\phi$.

\subsection{Fisher Information Analysis without A Priori Position Knowledge} \label{Sec:Eval_Analysis}

In this section, we consider the case in which {a priori} knowledge
of the agents' positions is unavailable, i.e., $\gp(\V{P})$ is
eliminated from \eqref{eq:FIM_Prior_Dis}. We first prove a general
theorem, describing the structure of the EFIM, followed by a special
case, where there is no {a priori} knowledge regarding the channel
parameters.

\begin{thm}\label{thm:EFIM_Coop}
When {a priori} knowledge of the agents' positions is unavailable,
and the channel parameters corresponding to different waveforms are
mutually independent, the EFIM for the agents' positions is a $2\Na
\times 2\Na$ matrix, structured as (\ref{eq:Coop_EFIM}) at the bottom of the page, \addtocounter{equation}{1}
where $\JA{\V{p}_k}$ and $\C_{kj}$ can be expressed in terms of the
RI:
    \begin{align*}
        \JA{\V{p}_k} & = \sum_{j\in \NB} \lambda_{kj} \, \R(\phi_{kj})
        \, , 
	\end{align*}
and
	\begin{align*}
        \C_{kj} & = \C_{jk} = \left(\lambda_{kj} + \lambda_{jk} \right) \, \R(\phi_{kj})
        \, ,
    \end{align*}
with $\lambda_{kj}$ given by \eqref{eq:Apd_RI_Inten} in Appendix \ref{apd:EFIM_Coop}. 
\end{thm}

\begin{IEEEproof}
See Appendix \ref{apd:EFIM_Coop}.
\end{IEEEproof}

\begin{rem} We make the following remarks.
\begin{itemize}
\item
To obtain the SPEB of a specific agent, we can apply EFI analysis
again and further reduce $\JE{\V{P}}$ into a $2 \times 2$ EFIM.

\item
The RI is the basic building block of the EFIM for localization, and each RI corresponds to an individual received waveform. The RII
$\lambda_{kj}$ is determined by the power and bandwidth of the
received waveform, the multipath propagation, as well as the {a priori} channel knowledge.
Note that each received waveform provides only one-dimensional
information for localization along the angle $\phi_{kj}$.

\item
The EFIM $\JE{\V{P}}$ can be decomposed into localization information from anchors and that from agents' cooperation. The former part is represented as a block-diagonal matrix whose non-zero elements are $\JA{\V{p}_k}$, for the $k$th agent, and each $\JA{\V{p}_k}$ is a weighted sum of RDMs over anchors. Hence the localization
information from anchors is not inter-related among agents. The latter part is a highly structured matrix consisting of RIs $\C_{kj}$. Hence the localization information from agents' cooperation is highly inter-related. This is intuitive since the effectiveness of the localization information provided by a particular agent depends on its position error.
\end{itemize}
\end{rem}


\begin{thm}\label{cor:RII_NoPrior}
When {a priori} knowledge of the agents' positions and the
channel parameters is unavailable, the EFIM for the agents'
positions is a $2\Na \times 2\Na$ matrix, structured as in
\eqref{eq:Coop_EFIM} shown at the bottom of the page, with
%
the RII $\lambda_{kj}$ given by
        \begin{align*}
        \lambda_{kj} = \begin{cases}
                {8\pi^2\beta^2}/{c^2} \,\cdot\, (1 - \chi_{kj})\, \mathsf{SNR}_{kj}^{(1)}
                \,, &  \; \text{LOS signal}, \\
                0 \,,  & \; \text{NLOS signal},
            \end{cases}
    \end{align*}
where $\beta$ is the effective bandwidth of transmitted waveform
$s(t)$
    \begin{align*}
        \beta = \left( \frac{\int_{-\infty}^{+\infty} f^2 \, |S(f)|^2 df}{
        \int_{-\infty}^{+\infty} |S(f)|^2 df}\right)^{1/2} ,
    \end{align*}
$\mathsf{SNR}_{kj}^{(1)}$ is the SNR of the \emph{first} path in
$r_{kj}(t)$:
    \begin{align}\label{eq:SNR_firstpath}
        \mathsf{SNR}_{kj}^{(1)} = \frac{{\big|}\A{kj}{1}{\big|}^2
        \int_{-\infty}^{+\infty}|S(f)|^2 df}{N_0}
        \,,
    \end{align}
and $0 \leq \chi_{kj} \leq 1$ is called the \emph{path-overlap coefficient}, which depends on the
first contiguous-cluster\footnote{The first contiguous-cluster is
the first group of non-disjoint paths. Two paths that arrive at time
$\tau_i$ and $\tau_j$ are called non-disjoint if $|\tau_i-\tau_j|$
is less than the duration of $s(t)$ \cite{SheWin:J10a}.} in LOS signals.
\end{thm}

\begin{IEEEproof}
See Appendix \ref{apd:EFIM_NoPrior}.
\end{IEEEproof}

\begin{rem} We make the following remarks.
\begin{itemize}
\item
The theorem shows that when {a priori} knowledge of channel
parameters is unavailable, the NLOS signals do not contribute to
localization accuracy, and hence these signals can be discarded. This agrees with the previous observations in \cite{JouDarWin:J08, QiKobSud:06, Mai:08} although the authors considered different models.

\item
For LOS signals, the RII is determined by the first
contiguous-cluster \cite{SheWin:J10a}, implying that it is
not necessary to process the latter multipath components.  In
particular, the RII is determined by the effective bandwidth
$\beta$, the first path's SNR, and the propagation effect characterized
by $\chi_{kj}$.

\item
Since $\chi_{kj} \geq 0$, path-overlap always
deteriorates the accuracy unless $\chi_{kj} = 0$, in which the first signal component $s(t-\tau^{(1)}_{kj})$ does not overlap with later components $s(t-\tau^{(l)}_{kj})$ for $l>1$.
\end{itemize}
\end{rem}

\begin{figure*}
	[!b] \vspace*{4pt} \hrulefill \normalsize \setcounter{MYtempeqncnt}{\value{equation}} \setcounter{equation}{14} 
    \begin{align}\label{eq:apd_RII_1}
        \V{R}_k(\V{r}_{kj}) & =
        \E_\V{P} \left\{ \Phi_{kj}\left(d_{kj},d_{kj}\right) \, \V{q}_{kj}\, \V{q}_{kj}^\text{T}
        \right\}  
        - \E_\V{P}\left\{\V{q}_{kj}\,\B\Phi_{kj}(d_{kj},\V{p}_k) \right\}\,
        \E_\V{P}\left\{\B\Phi_{kj}(\B{\kappa}_{kj},\B{\kappa}_{kj})\right\}^{-1}\,
        \E_\V{P}\left\{\B\Phi_{kj}(\V{p}_k,d_{kj})\,\V{q}_{kj}^\text{T}\right\}
    \end{align}
	\setcounter{equation}{\value{MYtempeqncnt}} \vspace*{-10pt} 
\end{figure*}

\begin{figure*}
	[!b] \vspace*{0pt} \hrulefill \normalsize \setcounter{MYtempeqncnt}{\value{equation}} \setcounter{equation}{16} 
	\begin{align}\label{eq:Coop_EFIM_Prior}
	\hspace{-4mm}
	    \JE{\V{P}} =  \Matrix{cccc}
	        {       \bJA{\V{p}_1} + \sum\limits_{j \in \NA\backslash\{1\}} \bar{\C}_{1,j}    &   -\bar{\C}_{1,2}  &   \cdots  &   -\bar{\C}_{1,\Na}  \\
	                -\bar{\C}_{1,2}            &   \bJA{\V{p}_2} + \sum\limits_{j \in \NA\backslash\{2\} } \bar{\C}_{2,j}  &           &   -\bar{\C}_{2,\Na}  \\
	                \vdots                  &                              &   \ddots  &      \\
	                -\bar{\C}_{1,\Na}          &   -\bar{\C}_{2,\Na}             &      &  \bJA{\V{p}_\Na} + \sum\limits_{j \in \NA\backslash\{\Na\}} \bar{\C}_{\Na,j} }
	    + \GP
	\end{align}
	\setcounter{equation}{\value{MYtempeqncnt}} \vspace*{-6pt} 
\end{figure*}

\subsection{Fisher Information Analysis with A Priori Position Knowledge} \label{Sec:Eval_PrioriPosition_Knowledge}

We now consider the case in which the {a priori} knowledge of the agents' positions, characterized by $\gp(\V{P})$, is available. We first derive the EFIM, based on which we prove that agents and anchors can be treated in a unified way under this framework. We then present a special scenario in which the a priori knowledge of the agents' positions satisfies certain conditions so that we can gain insights into the EFIM.

\begin{thm}\label{thm:EFIM_Prior}
When a priori knowledge of the agents' positions is available, and
the channel parameters corresponding to different waveforms are
mutually independent, the EFIM for the agents' positions is a $2\Na
\times 2\Na$ matrix, given by\footnote{Note that $\JE{\V{P}}$ in
\eqref{eq:apd_EFIM_Prior} does not depend on any particular value
of the random vector $\V{P}$, whereas $\JE{\V{P}}$ in
\eqref{eq:Coop_EFIM} is a function of the deterministic vector
$\V{P}$.}
    \begin{align}\label{eq:apd_EFIM_Prior}
        \JE{\V{P}}  & = \JA{\V{P}} + \JC{\V{P}} + \B{\Xi}_{\V{P}}  ,
    \end{align}
where
    \begin{align*}
        \left[\, \JA{\V{P}} \,\right]_{2k-1:2k,2m-1:2m}
        & =  \begin{cases}
                \sum_{j\in\NB} \V{R}_k(\V{r}_{kj})\, ,  & k = m\,,\\
                \V{0}  \, , & k \neq m\,,
              \end{cases} 
	\end{align*}
	\begin{align*}
        & \hspace{-6mm} \left[\, \JC{\V{P}} \,\right]_{2k-1:2k,2m-1:2m} \\
        & = \begin{cases}
                \sum_{j\in\NA\backslash\{k\}}
                \left[\, \V{R}_k(\V{r}_{kj})+\V{R}_k(\V{r}_{jk}) \,\right] \, ,  &  k = m\,,\\
                - [\, \V{R}_k(\V{r}_{km}) + \V{R}_k(\V{r}_{mk}) \,]
                \, , &   k \neq m\,,
            \end{cases} 
		\end{align*}
and
		\begin{align*}
        \GP & = \E_{\V{P}}\!\left\{
        - \frac{\partial^2}{\partial \V{P} \partial \V{P}^\text{T}}
        \ln \gp(\V{P}) \right\}
        \, , 
    \end{align*}
with $\V{R}_k(\V{r}_{kj})\in\mathbb{R}^{2\times2}$ given by (\ref{eq:apd_RII_1}) shown at the bottom of the page. \addtocounter{equation}{1}
Block matrix $\bm{\Phi}_{kj}\left(\V{x},\V{y}\right)$ in (\ref{eq:apd_RII_1}) is defined as \eqref{eq:apd_def_phi_xy} in Appendix \ref{apd:EFIM_Coop}.
\end{thm}

\begin{IEEEproof}
See Appendix \ref{sec:Apd_coop_Prior}.
\end{IEEEproof}

\begin{rem}
The EFIM for agents' positions is derived in (\ref{eq:apd_EFIM_Prior}) for the case when a priori knowledge of the agents' positions is available. Compared to (\ref{eq:Coop_EFIM}) in the Theorem \ref{thm:EFIM_Coop}, the EFIM in (\ref{eq:apd_EFIM_Prior}) retains the same structure of the localization information from both anchors and cooperation, except that all RIs in Theorem \ref{thm:EFIM_Prior} are obtained by averaging the $2\times 2$ matrices over the possible agents' positions. In addition, the localization information from the position knowledge is characterized in terms of an additive component $\B{\Xi}_\V{P}$. This knowledge improves localization because $\B{\Xi}_\V{P}$ is positive semi-definite.
\end{rem}

Based on the result of Theorem \ref{thm:EFIM_Prior}, we can now treat anchors and agents in a unified way, as will be shown in the following theorem.

\begin{thm}\label{cor:anc_agent}
Anchors are equivalent to agents with infinite {a priori} position
knowledge in the following sense: when the $k$th agent has infinite
a priori position knowledge, i.e., $\bm\Xi_{\mathbf{p}_{k}} =
\lim_{t^2\rightarrow\infty}\Diag{t^2,t^2}$, then
    \begin{align*}
        \JE{\V{P}_{\bar{k}}} =
        \left[\, \JE{\V{P}}\, \right]_{\bar{k}}
        \, ,
    \end{align*}
where $\V{P}_{\bar{k}}$ is the vector $\V{P}$ without rows $2k-1$ to
$2k$, and $\left[\, \JE{\V{P}}\, \right]_{\bar{k}}$ is the matrix
$\JE{\V{P}}$ without rows $2k-1$ to $2k$ and columns $2k-1$ to $2k$.
\end{thm}

\begin{IEEEproof}
See Appendix \ref{sec:Apd_anc_agent}.
\end{IEEEproof}

\begin{rem}
The theorem shows mathematically that agents are equivalent to anchors if they have infinite {a priori} position knowledge, which agrees with our intuition. As such, it is not necessary to distinguish between agents and anchors. This view will facilitate the analysis of location-aware networks and the design of localization algorithms: every agent can treat the information coming from anchors and other cooperating agents in a unified way.
\end{rem}

The general expression of the EFIM for the case with a priori position knowledge is given in \eqref{eq:apd_EFIM_Prior}, which is much more involved than that for the case without position knowledge in \eqref{eq:Coop_EFIM}. However, in the special case when
    \begin{align}\label{eq:conditionPrior}
        \E_{\V{P}} \left\{ g( \V{P}) \right\}
        = g( \E_{\V{P}} \left\{ \V{P} \right\})
        \,,
    \end{align}
for the functions $g(\cdot)$ involved in the derivation of the EFIM
(see Appendix \ref{sec:Apd_coop_Prior}),\footnote{ This occurs when
every agent's {a priori} position distribution is concentrated in a
small area relative to the distance between the agent and the other
nodes, so that $g(\V{P})$ is flat in that area.} we can gain
insight into the structure of the EFIM as shown by the following
corollary.


\begin{cor}\label{thm:EFIM_Coop_Prior}
When the {a priori} distribution of the agents' positions satisfies \eqref{eq:conditionPrior}, and the channel parameters corresponding to different waveforms are mutually independent, the EFIM for the agents' positions is a $2\Na \times 2\Na$ matrix, structured as (\ref{eq:Coop_EFIM_Prior}) shown at the bottom of the page, \addtocounter{equation}{1}
where $\bJA{\V{p}_k}$ and $\bar{\C}_{kj}$ can be expressed in terms
of the RI:
    \begin{align*}
        \bJA{\V{p}_k} & =
        \sum_{j\in \NB} \bar{\lambda}_{kj} \, \R(\bar{\phi}_{kj})
        \, , \\
        \noalign{\noindent and \vspace{\jot}}
        \bar{\C}_{kj} & = \bar{\C}_{jk} =
        \left(\bar{\lambda}_{kj} + \bar{\lambda}_{jk} \right) \, \R(\bar{\phi}_{kj})
        \, ,
    \end{align*}
where $\bar{\V{P}} = \E_{\V{P}} \left\{ \V{P} \right\}$,
$\bar\lambda_{kj}$ is the RII given in \eqref{eq:Apd_RI_Inten}
evaluated at $\bar{\V{P}}$, and $\bar\phi_{kj}$ is the angle from
$\bar{\V{p}}_k$ to $\bar{\V{p}}_j$.
\end{cor}

\begin{IEEEproof}
See Appendix \ref{sec:Apd_coop_Prior}.
\end{IEEEproof}

 
\begin{figure*}
	[!b] \vspace*{4pt} \hrulefill \normalsize \setcounter{MYtempeqncnt}{\value{equation}} \setcounter{equation}{17} 
	\begin{align}\label{eq:Jc_tracking}
        \JC{\V{P}} & =
        \Matrix{ccccc}
            {       \C_{1,2}   &   -\C_{1,2}       &    &    \\
                    -\C_{1,2}                   &  \C_{1,2} + \C_{2,3}  &   -\C_{2,3}  &    \\
                    &   -\C_{2,3}       &   \ddots         &  \ddots   &    \\
                    &  &    \ddots &   \C_{N-2,N-1} + \C_{N-1,N}     &   -\C_{N-1,N}   \\
                    &   &     &   - \C_{N-1,N}   &  \C_{N-1,N} }
    \end{align}
	\setcounter{equation}{\value{MYtempeqncnt}} \vspace*{-10pt} 
\end{figure*}

\begin{figure*}
	[!b] \vspace*{0pt} \hrulefill \normalsize \setcounter{MYtempeqncnt}{\value{equation}} \setcounter{equation}{18} 
	\begin{align}\label{eq:Eval_EFIM_Cons}
	    \JE{\V{P}_{\!n+1}}
	    = \left[ \begin{array}{cc} \JE{\V{P}_{\!n}} + \V{M}_{n,n+1}      &   -\V{M}_{n,n+1} \, \V{K}_n \\
		                      -\V{K}_n^\text{T} \, \V{M}_{n,n+1}                 &  \V{J}_{\text{A},n+1} + \V{K}_n^\text{T} \, \V{M}_{n,n+1} \, \V{K}_n\end{array}\right]
	\end{align}
	\setcounter{equation}{\value{MYtempeqncnt}} \vspace*{-6pt} 
\end{figure*}

\subsection{Discussions}

We will now discuss the results derived in the previous sections. Our discussion includes 1) the EFIM for the agents in non-cooperative localization, 2) an application of the cooperative localization to tracking, 3) a recursive method to construct an EFIM for large networks, and 4) the extension to three-dimensional scenarios.

\subsubsection{Non-Cooperative Localization}
When the agents do not cooperate, the matrices corresponding to the agents' cooperation in \eqref{eq:Coop_EFIM} in Theorem \ref{thm:EFIM_Coop} and \eqref{eq:Coop_EFIM_Prior} in Corollary \ref{thm:EFIM_Coop_Prior} are discarded. In particular, the EFIM $\JE{\V{P}}$ in Theorem \ref{thm:EFIM_Coop} reverts to
    \begin{align*}
        \JE{\V{P}} = \Diag{ \JA{\V{p}_1}\, , \JA{\V{p}_2}\, , \,\ldots \, , \JA{\V{p}_\Na}} ,
    \end{align*}
and hence the $2 \times 2$ EFIM for the $\Th{k}$ agent is equal to
$\JE{\V{p}_k} = \JA{\V{p}_k} $.
Similarly, the EFIM $\JE{\V{P}}$ in Corollary \ref{thm:EFIM_Coop_Prior} reverts to
    \begin{align*}
        \JE{\V{P}} = \Diag{ \bJA{\V{p}_1}\, , \bJA{\V{p}_2}\, , \, \ldots \, , \bJA{\V{p}_\Na}} + \GP  \, .
    \end{align*}
Furthermore, when the agents' positions are independent {a priori},
$\GP = \Diag{\bm\Xi_{\V{p}_1}, \bm\Xi_{\V{p}_2}, \ldots, \bm\Xi_{\V{p}_\Na}}$ and the
$2 \times 2$ EFIM for the $\Th{k}$ agent can be written as
$\JE{\V{p}_k} = \bJA{\V{p}_k} + \bm\Xi_{\V{p}_k}$.

\subsubsection{Spatial vs. Temporal Cooperation for Localization}
Rather than multiple agents in cooperation, a single agent can
``cooperate'' with itself over time. Such temporal cooperative
localization can easily be analyzed within our framework, as
follows.

Consider a single agent moving in sequence to $N$ different
positions according to piecewise linear walk and receiving
waveforms from neighboring anchors at each position. The $N$ positions can be written as $\V{P} =
\left[\,\V{p}_1^\text{T} \;\; \V{p}_2^\text{T} \;\; \cdots \;\;
\V{p}_N^\text{T}\,\right]^\text{T}$, and we can consider the
scenario as $N$ agents in cooperation. The likelihood of the
observation is
    \begin{align*}
        f\left( \V{r},\hat{\V{d}}|\B\Theta \right)
        =   \prod_{k=1}^N \prod_{j\in\NB} f\left( \V{r}_{kj} | \V{p}_k, \V{p}_j\right)
            \, \prod_{k=1}^{N-1} f\left( \hat{d}_k | \V{p}_k, \V{p}_{k+1}\right) ,
    \end{align*}
where $\hat{\V{d}} = {\big[}\, \hat{d}_1\; \hat{d}_2\; \cdots \; \hat{d}_{N-1} \,{\big]}^\text{T}$ in which $\hat{d}_k$ is the measurement of the distance $d_k=\|\V{p}_k-\V{p}_{k+1}\|$ between $\V{p}_k$ and $\V{p}_{k+1}$.\footnote{We assume that the agent has other navigation devices, such as inertial measurement unit (IMU), odometer, or pedometer, to measure the distance between positions.} By applying Theorem \ref{thm:EFIM_Coop}, we have the EFIM for $\V{P}$ as $\JE{\V{P}} = \JA{\V{P}} + \JC{\V{P}}$ where
    \begin{align*}
        \JA{\V{P}} & = \Diag{\JA{\V{p}_1},\JA{\V{p}_2},\, \ldots, \JA{\V{p}_N}} ,
	\end{align*}
and $\JC{\V{P}}$ is given by (\ref{eq:Jc_tracking}) shown at the bottom of the page, \addtocounter{equation}{1}
in which $\C_{k,k+1} = \nu_k \, \R(\phi_{k,k+1})$ with
$\phi_{k,k+1}$ denoting the angle from $\V{p}_k$ to $\V{p}_{k+1}$ and
    \begin{align*}
        \nu_k = \E_{\hat{\V{d}}} \left\{
        -\frac{\partial^2}{\partial d_{k}^2} \ln f\left( \hat{d}_k |
        \V{p}_k, \V{p}_{k+1}\right)\right\}.
    \end{align*}
By further applying the notion of EFI, we can obtain the EFIM $\JE{\V{p}_k}$ for each position $\V{p}_k$. Note that this analysis can be extended to cooperation among multiple mobile agents over time, so that both cooperation over space and time are explored simultaneously.

\subsubsection{Recursive Formula for EFIM}

The structure of the EFIM in \eqref{eq:Coop_EFIM} and \eqref{eq:Coop_EFIM_Prior} enables us to extend the EFIM when agents join or leave the cooperative network. We will develop a recursive formula to construct the EFIM in the following.

Consider a network with $n$ agents  in cooperation without a priori knowledge of their positions, and the EFIM for agents' positions
$\JE{\V{P}_{\!n}}$ where $\V{P}_{\!n} = [\, \V{p}_1^\text{T}\; \cdots \;\V{p}_n^\text{T}\,]^\text{T}$ can be obtained by \eqref{eq:Coop_EFIM}.  If a new agent enters the cooperative network, then the EFIM for the $n+1$ agents is given by (\ref{eq:Eval_EFIM_Cons}), shown at the bottom of the page,\addtocounter{equation}{1}
where $\V{J}_{\text{A},n+1}$ is the EFIM for the $\Th{(n\!+\!1)}$ agent corresponding to the localization information from anchors, $\V{M}_{n,n+1}$ is the localization information from the cooperation between the $\Th{(n+1)}$ agent and the other $n$ agents, given by
    \begin{align*}
        \V{M}_{n,n+1} = \Diag{\C_{1,n+1}\, , \,
        \C_{2,n+1}\, , \, \ldots \, , \, \C_{n,n+1}} ,
    \end{align*}
and $\V{K}_n\in\mathbb{R}^{2n\times2}$ is given by
    \begin{align*}
        \V{K}_n = {\Matrix{ccccccc}
                {\V{I}_{2\times2}  &   \V{I}_{2\times2}    &   \cdots  &  \V{I}_{2\times2} }}^\text{T} .
    \end{align*}
Note that when the {a priori} knowledge of the agents' positions is available, we need to consider the contribution of $\GP$, and the EFIM for the $n+1$ agents can be constructed in a similar way.

Similarly, when a certain agent, say $k$, leaves the network, we need to eliminate rows $2k-1$ to $2k$ and columns $2k-1$ to $2k$ in $\JE{\V{P}_{\!n}}$, as well as subtract all corresponding $\V{C}_{kj}$ for $j \in \NA \backslash \{k\}$ from the diagonal of $\JE{\V{P}_{\!n}}$.

\subsubsection{Extension to 3D Localization}

All the results obtained thus far can be easily extended to the three-dimensional scenario, in which $\V{p}_k = [\, x_k \; y_k \; z_k \, ]^\text{T}$. The SPEB of the $k$th agent is defined as $\PEB(\V{p}_k)=\left[ \JTH^{-1} \right]_{3 \times 3,k}$. Following the steps leading to \eqref{eq:Coop_EFIM} and \eqref{eq:Coop_EFIM_Prior}, we can obtain a corresponding $3\Na \times 3\Na$ EFIM involving the RDMs $\R(\varphi_{kj},\phi_{kj})$ for $k\in\NA$ and $j\in\NB\cup\NA$, where
    \begin{align*}
        \R(\varphi,\phi) \triangleq \q \, \q^\text{T}
        \, ,
    \end{align*}
with $\varphi$ and $\phi$ denoting the angles in the spherical coordinates, and $\q = \left[\,\cos\varphi \cos\phi \;\; \sin\varphi \cos\phi  \;\; \sin\phi \,\right]^\text{T}$.

%
%
\begin{figure}[t]
    \psfrag{x}[l][][1.2]{\hspace{0mm}$x$}
    \psfrag{x1}[l][][1.2]{\hspace{-3mm}$x^*$}
    \psfrag{y}[l][][1.2]{\hspace{-1mm}$y$}
    \psfrag{y1}[l][][1.2]{\hspace{-2mm}$y^*$}
    \psfrag{mu1N}[l][][1.2]{\hspace{-5mm}$\sqrt{\mu}$}
    \psfrag{mu2N}[l][][1.2]{\hspace{-5mm}$\sqrt{\eta}$}
    \psfrag{beta}[l][][1.2]{\hspace{-2mm}$\vartheta$}
    \begin{center}
	\includegraphics[angle=0,width=0.8\linewidth,draft=false]{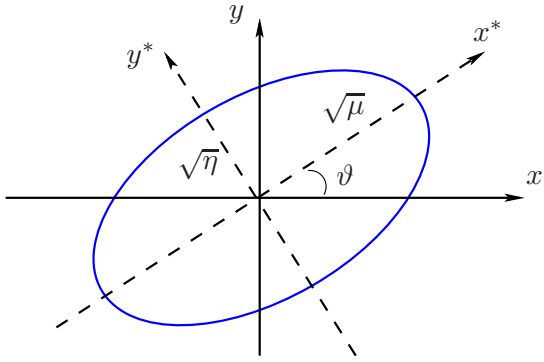}
    \caption{\label{fig:ellipse_1}
    Geometric interpretation of the EFIM as an information ellipse.
    In the rotated coordinate system (rotated
    over an angle $\vartheta$), the major and minor axes of the ellipse are given by $\sqrt{\mu}$ and $\sqrt{\eta}$, respectively.}
    \end{center}
\end{figure}

\section{Geometric Interpretation of EFIM for
Localization}\label{Sec:Geom}

In this section, we present a geometric interpretation of the EFIM
for localization. 
This interpretation not only provides insights into the essence of
localization problems, but also facilitates the analysis of localization
systems, design of localization algorithms, and deployment of
location-aware networks. We begin with the non-cooperative case,
and then extend to the cooperative case. Based on these results, we
derive scaling laws of the SPEB for both non-cooperative and
cooperative location-aware networks.

\subsection{Interpretation for Non-Cooperative Localization}\label{Sec:Geom_Eigen}

When an agent only communicates with neighboring anchors, the EFIM can be written as\footnote{To simplify the notation, we will suppress the agent's index in the subscript.}
    \begin{align}\label{eq:Anal_EFIM}
        \JE{\V{p}} = \sum_{j\in\NB} \lambda_{j} \, \R(\phi_j) 
            \triangleq  \V{U}_{\vartheta} \Matrix{cc} {\mu  &  0 \\ 0  &
            \eta }  \V{U}_{\vartheta}^\text{T}
        \, ,
    \end{align}
where $\mu$ and $\eta$ are the eigenvalues of $\JE{\V{p}}$, with $\mu \geq \eta$,
and $\V{U}_{\vartheta}$ is a rotation matrix with angle
${\vartheta}$, given by
    \begin{align*}
        \V{U}_{\vartheta}
        = \Matrix{cc} { \cos{\vartheta}       &    -\sin {\vartheta} \\
                        \sin{\vartheta}       &    \cos {\vartheta} } .
    \end{align*}
%
The first and second columns of $\V{U}_{\vartheta}$ are the
eigenvectors corresponding to eigenvalues $\mu$ and $\eta$,
respectively. By the properties of eigenvalues, we have
    \begin{align*}
        \mu + \eta = \T\left\{ \JE{\V{p}}\right\} =  \sum_{j \in \NB} \lambda_{j}
        \, .
    \end{align*}
Note in \eqref{eq:Anal_EFIM} that $\JE{\V{p}}$ depends only on
$\mu$, $\eta$, and $\vartheta$, and we will denote $\JE{\V{p}}$ by
$\V{F}(\mu, \eta, \vartheta)$ when needed.


\begin{pro}\label{pro:SPEB_CIndep}
The SPEB is  independent of the  coordinate system.
\end{pro}

\begin{IEEEproof}
See Appendix \ref{Sec:Apd_pro}.
\end{IEEEproof}

\begin{rem}
The proposition implies that if we rotate the original coordinate
system by an angle $\vartheta$ prescribed by \eqref{eq:Anal_EFIM} and
denote the agent's position in the new coordinate by $\V{p}^*$, then
the SPEB is
    \begin{align*}
        \PEB(\V{p}) = \PEB(\V{p}^*)
        = \T\left\{ \Matrix{cc} {\mu  &  0 \\ 0  &
        \eta    }^{-1}  \right\}
        = \frac{1}{\mu} + \frac{1}{\eta}
        \, .
    \end{align*}
The EFIM in the new coordinate system is diagonal, and thus the localization information in these new axes is decoupled. Consequently, the SPEB is also decoupled in these two orthogonal directions.
\end{rem}

\begin{defi}[Information Ellipse]
Let $\V{J}$ be a $2 \times 2$ positive definite matrix. The information
ellipse of $\V{J}$ is defined as the sets of points $\V{x} \in
\mathbb{R}^2$ such that
    \begin{align*}
        \V{x} ~ \V{J}^{-1} \V{x}^\text{T} = 1
        \, .
    \end{align*}
\end{defi}

Geometrically, the EFIM in \eqref{eq:Anal_EFIM} corresponds to an
information ellipse with major and minor axes equal to $\sqrt{\mu}$ and $\sqrt{\eta}$, respectively, and a rotation $\vartheta$ from the
reference coordinate, as depicted in Fig.~\ref{fig:ellipse_1}.
Hence, the information ellipse is completely characterized by
$\mu$, $\eta$, and $\vartheta$.
%
Note that the RI is expressed as $\lambda \,\R(\phi) = \V{F}(\lambda, 0, \phi)$, and it corresponds to a degenerate ellipse. In the following proposition, we will show how an anchor contributes to the
information ellipse of an agent.

\begin{figure}[t]
    \psfrag{x}[][][1.2]{\hspace{1mm}$x$}
    \psfrag{y}[][][1.2]{$y$}
    \psfrag{mu1}[][][1.1]{\hspace{-2mm}$\sqrt{\mu}$}
    \psfrag{mu2}[][][1.1]{\hspace{-2mm}$\sqrt{\eta}$}
    \psfrag{mu1N}[][][1.1]{ \hspace{-3mm}$\sqrt{\tilde\mu}$}
    \psfrag{mu2N}[][][1.1]{\hspace{-5mm}$\sqrt{\tilde\eta}$}
    \psfrag{beta}[][][1]{\hspace{-4mm}  $\tilde\vartheta$}
    \psfrag{alpha}[][][1]{\hspace{-6mm} $\phi'$}
    \psfrag{nu}[][][1.1]{\hspace{-1mm}$\nu$}
    \begin{center}
    \includegraphics[angle=0,width=0.8\linewidth,draft=false]{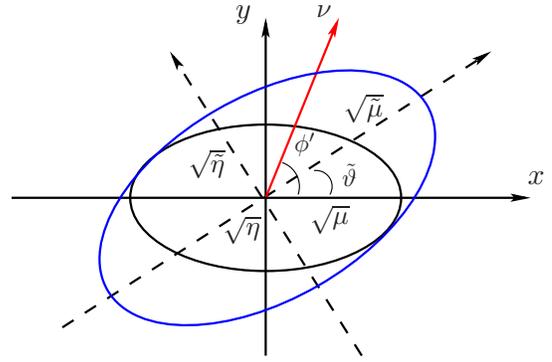}
    \caption{\label{fig:ellipse_2}
    Updating of the information ellipse for non-cooperative localization. The original
    information ellipse of the agent is characterized by $\V{F}(\mu,\eta,0)$. The RI from an additional anchor
    is given by $\V{F}(\nu,0,\phi)$. The new information ellipse of the agent then grows along
    the direction $\phi'$, but not along the orthogonal direction. The new  information ellipse
    corresponds
    to $\V{F}(\tilde\mu,\tilde\eta,\tilde\vartheta$). }
    \end{center}
\end{figure}


\begin{pro}
Let $\JE{\V{p}} = \V{F}(\mu, \eta, \vartheta)$ and $\PEB(\V{p})$ denote the EFIM and the SPEB of an agent, respectively. When that agent obtains RI $\V{F}(\nu, 0, \phi)$ from a new anchor, the new EFIM for the agent will be
    \begin{align*}
        \tilde{\V{J}}_\text{e}({\V{p}})
        & = \V{F}(\tilde{\mu}, \tilde\eta, \tilde{\vartheta})
        \nonumber \\
        & = \V{F}(\mu, \eta, \vartheta) + \V{F}(\nu, 0, \phi)
        \, ,
    \end{align*}
where the parameters for the new information ellipse are
    \begin{align*}
        \tilde\mu & =  \frac{\mu + \eta +
            \nu}{2} + \frac{1}{2} \sqrt{\left[ \mu - \eta + \nu \cos 2 \phi' \right]^2 + \nu^2\sin^2 2\phi'} \\
        \tilde\eta & =  \frac{\mu + \eta +
            \nu}{2} - \frac{1}{2} \sqrt{\left[ \mu - \eta + \nu \cos 2 \phi' \right]^2 + \nu^2 \sin^2 2\phi'}\, \\
        \noalign{\noindent and \vspace{\jot}}
        \tilde\vartheta  & = \vartheta +
            \frac{1}{2}\arctan \frac{\nu \sin 2\phi'}{\mu - \eta + \nu \cos 2 \phi'}
        \,,
    \end{align*}
with $\phi' \triangleq \phi - \vartheta$. Correspondingly, the new
SPEB becomes
    \begin{align}\label{eq:Anal_CRLB_COOP}
        \tilde{\PEB}(\V{p}) =
        \frac{1}{\tilde\mu} + \frac{1}{\tilde\eta}
        = \frac{\mu + \eta + \nu}{\mu\eta +
        \nu \left[ \eta + (\mu - \eta)\sin^2\phi'\right]}
        \, .
    \end{align}
\end{pro}

\begin{rem}
The geometric interpretation for the proposition is depicted in Fig.~\ref{fig:ellipse_2}. For a fixed RII $\nu$, we see from \eqref{eq:Anal_CRLB_COOP} that $\tilde{\PEB}(\V{p})$ can be minimized through $\phi'$ (equivalently, through $\phi$) in the denominator:
    \begin{align*}
        \min_{\phi}\; \tilde{\PEB}(\V{p}) = \frac{\mu + \eta + \nu}{\mu(\eta + \nu)}
        \, ,
    \end{align*}
and the minimum is achieved when $\phi = \vartheta \pm {\pi}/{2}$.
In such a case, the anchor is along the direction of the eigenvector corresponding to the smallest eigenvalue $\eta$. Observe also that the denominator in \eqref{eq:Anal_CRLB_COOP} is equal to $\tilde\mu \cdot \tilde\eta$, which is proportional to the squared \emph{area} of the new information ellipse corresponding to $\tilde{\V{J}}_\text{e}({\V{p}})$. Hence, for a fixed $\nu$, the minimum SPEB is achieved when the new anchor is along the minor axis of the information ellipse corresponding to $\JE{\V{p}}$. Equivalently, this choice of anchor position maximizes the area of the new information ellipse. 

On the other hand, the maximum SPEB occurs when the anchor is along the direction of the eigenvector corresponding to the largest eigenvalue $\mu$, i.e., the major axis of the information ellipse corresponding to $\JE{\V{p}}$. Equivalently, this minimizes the area of the new information ellipse, and thus
    \begin{align*}
        \max_{\phi}\; \tilde{\PEB}(\V{p}) = \frac{\mu + \eta + \nu}{\eta(\mu + \nu)}
        \, ,
    \end{align*}
and the maximum is achieved when $\phi = \vartheta \pm \pi$. Note
also that
    \begin{align*}
        \frac{1}{\mu}<
        \tilde{\PEB}(\V{p}) \leq  \PEB(\V{p})
        \, ,
    \end{align*}
where the left-hand side $1/{\mu} = \lim_{\nu \rightarrow \infty} \min_{\phi}\; \tilde{\PEB}(\V{p})$, and the right-hand side $ \PEB(\V{p}) = \lim_{\nu \rightarrow 0} \tilde{\PEB}(\V{p})$.

\end{rem}

\subsection{Interpretation for Cooperative Localization}

The EFIM for all the agents in cooperative location-aware network is given respectively by \eqref{eq:Coop_EFIM_Prior} and \eqref{eq:Coop_EFIM} for the cases with and without a priori position knowledge. Further applying the notion of EFI, one can obtain the EFIM for individual agents. In general, the exact EFIM expression for the individual agents is complicated. However, we can find lower and upper bounds on the individual EFIM to gain some insights into the localization problem.


\begin{pro}\label{pro:MultiAnt}
Let $\JA{\V{p}_k} = \V{F}(\mu_k,\eta_k,\vartheta_k)$ denote the EFIM for agent $k$ that corresponds to the localization information from anchors, and let $\C_{kj}=\V{F}(\nu_{kj},0,\phi_{kj})$ denote the RI for that agent obtained from cooperation with agent $j$. The EFIM $\JE{\V{p}_k}$ for agent $k$ can be bounded as follows:
    \begin{align*}
        \JL{\V{p}_k}  \preceq  \JE{\V{p}_k} \preceq
        \JU{\V{p}_k}   \, ,
    \end{align*}
where
    \begin{align}\label{eq:Anal_Lower}
        \JL{\V{p}_k}  & = \JA{\V{p}_k}
        + \sum_{j\in\NA\backslash\{k\}} \xi_{kj}^\text{L} \, \C_{kj}
        \, , 
	\end{align}
	\begin{align}
        \JU{\V{p}_k} & = \JA{\V{p}_k}
        + \sum_{j\in\NA\backslash\{k\}} \xi_{kj}^\text{U} \, \C_{kj}
        \, , \label{eq:Anal_Upper}
    \end{align}
with coefficients $0 \leq \xi_{kj}^\text{L} \leq \xi_{kj}^\text{U}
\leq 1$ given by \eqref{eq:apd_chi_L} and \eqref{eq:apd_chi_U}.
\end{pro}

\begin{IEEEproof}
See Appendix \ref{Sec:Apd_pro}.
\end{IEEEproof}

\begin{rem}
The bounds for the EFIM can be written as weighted sums of RIs from
the neighboring nodes, and such linear forms can facilitate
analysis and design of location-aware networks. Moreover, it turns
out that $\xi_{kj}^\text{L}=\xi_{kj}^\text{U}$ when there are only
two agents in cooperation, leading to the following corollary.
\end{rem}

\begin{figure}[t]
    \psfrag{A1}[][][1.2]{Agent $1$}
    \psfrag{A2}[][][1.2]{Agent $2$}
    \psfrag{C12}[][][1.2]{$\C_{1,2}$}
    \psfrag{JA1}[][][1.2]{\hspace{3mm} $\JA{\V{p}_1}$}
    \psfrag{JE1}[][][1.2]{$\JE{\V{p}_1}$}
    \psfrag{JA2}[][][1.2]{\hspace{2mm} $\JA{\V{p}_2}$} 
	\psfrag{nu}[][][1.2]{$\nu$}
    \psfrag{alpha}[][][1.2]{\hspace{-2mm}$\phi_{1,2}$}
    \begin{center}
    \includegraphics[angle=0,width=0.8\linewidth,draft=false]{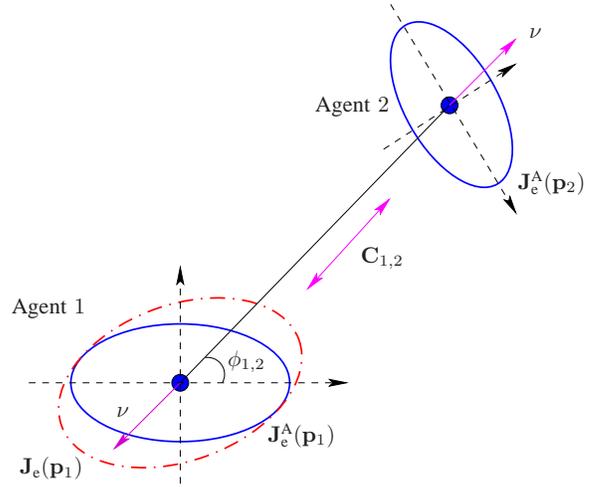}

\caption{Updating of the information ellipse for cooperative
localization. Based on the anchors, the $k$th agent has information
$\JA{\V{p}_k}$. The cooperative information between the two agents
is given by $\C_{1,2}=\V{F}(\nu,0,\phi_{1,2})$. The total EFIM for
agent 1 is then $\JE{\V{p}_1}=\JA{\V{p}_1} + \xi_{1,2} \C_{1,2}$.
The new information ellipse grows along the line connecting the two
agents. \label{fig:coop_1}}
	\end{center}
\end{figure}

%
%
%

%

\begin{cor}\label{cor:SPEB_TwoAgent}
Let $\JA{\V{p}_1} = \V{F}(\mu_1, \eta_1, \vartheta_1)$ and $\JA{\V{p}_2} = \V{F}(\mu_2, \eta_2, \vartheta_2)$ denote the EFIMs for agent 1 and 2
from anchors, respectively, and let $\C_{1,2} = \V{F}(\nu_{1,2}, 0, \phi_{1,2})$ denote the RI from their cooperation. The EFIMs for the two agents are given, respectively, by (see also Fig.~\ref{fig:coop_1})
    \begin{align*}
        \JE{\V{p}_1} & = \JA{\V{p}_1} + \xi_{1,2} \, \nu_{1,2}\, \R(\phi_{1,2})\,, \\
        \noalign{\noindent and \vspace{\jot}}
        \JE{\V{p}_2} & = \JA{\V{p}_2} + \xi_{2,1} \, \nu_{1,2}\, \R(\phi_{1,2})
        \, ,
    \end{align*}
where 
\begin{align*}
	\xi_{1,2} & = \frac{1}{{1+\nu_{1,2}\,\Delta_2(\phi_{1,2})}}\,, \\
    \noalign{\noindent and \vspace{\jot}}
	\xi_{2,1} &= \frac{1}{{1 + \nu_{1,2} \, \Delta_1(\phi_{1,2})}}\,,
\end{align*}
with
    \begin{align*}
        \Delta_k(\phi_{1,2}) = \V{q}_{12}^\text{T} \left[\,\JA{\V{p}_k}\,\right]^{-1} \V{q}_{12}\,, 
    \end{align*}
for $k=1,2$.
\end{cor}

\begin{rem} The results follow directly from Proposition \ref{pro:MultiAnt}. We make the following remarks.
\begin{itemize}
\item
Cooperation provides agent 1 with RI $\xi_{1,2} \, \nu_{1,2}\,
\R(\phi_{1,2})$ with $0\leq \xi_{1,2} \leq 1$.
Hence agent 1 obtains a RII $\xi_{1,2} \, \nu_{1,2}$ from cooperation instead of the full RII $\nu_{1,2}$. This degradation in RII is due to the inherent uncertainty of the second agent's position. We introduce the \emph{effective} RII $\tilde{\nu}_{1,2} = \xi_{1,2} \,
\nu_{1,2}$.

\item
The effective RII has the following geometric interpretation.
The value $\Delta_2(\phi_{1,2})$ is the DPEB of agent 2 (based solely on the anchors) along the angle $\phi_{1,2}$ between the two agents. This implies that the larger the uncertainty of agent 2 along the angle $\phi_{1,2}$, the less effective cooperation is. For a given $\Delta_2(\phi_{1,2})$, the effective RII $\tilde{\nu}_{1,2}$ increases monotonically with $\nu_{1,2}$, and has the following asymptotic limits:
    \begin{align*}
        \lim_{\nu_{1,2} \rightarrow 0} \tilde\nu_{1,2} &= 0
        \, , \nonumber \\
        \lim_{\nu_{1,2} \rightarrow \infty} \tilde\nu_{1,2} &=
        1 / \Delta_2(\phi_{1,2})
        \, .
    \end{align*}
Hence the maximum effective RII that agent 2 can provide to agent
1 equals the inverse of the DPEB of agent 2 (based solely on the
anchors) along the angle $\phi_{1,2}$ between the two agents.

\item
When i) the two agents happen to be oriented such that $\phi_{1,2} =
\vartheta_2$, and ii) agent 2 is certain about its position along
that angle ($\mu_2 = +\infty$), then $\Delta_2(\phi_{1,2})=0$ and 
$\JE{\V{p}_1} = \JA{\V{p}_1} + \C_{1,2}$, i.e., agent 2 can be
thought of as an anchor from the standpoint of providing RI to agent
1. From this perspective, anchors and agents are equivalent for
localization, where anchors are special agents with zero SPEB, or equivalently, infinite $\JA{\V{p}_k}$ in all directions.
\end{itemize}
\end{rem}

\subsection{Scaling Laws for Location-Aware Networks}\label{sec:scaling_law}

In this section, we derive scaling laws of the SPEB for both non-cooperative and cooperative location-aware networks. Scaling laws give us insight into the benefit of cooperation for localization in large networks. As we will see, agents and anchors contribute equally to the scaling laws for cooperative location-aware networks.

We focus on two types of random networks: \emph{dense} networks and
\emph{extended} networks \cite{GupKum:00, OzgLevTse:07}. In both types of networks, we consider the $\Nb$ anchors and $\Na$ agents randomly located (uniformly distributed) in the plane. In dense networks, adding nodes increases the node density, while the area remains
constant. In extended networks, the area increases proportional to the number of nodes, while both the anchor and the agent densities remain constant. 
Without loss of generality, we consider one round of transmission from each node to another. All transmission powers are the same, while large- and small-scale fading can be arbitrary. Medium access control is assumed so that these signals do not interfere with one another.


\begin{defi}[Scaling of SPEB]
Consider a network with $n$ nodes randomly located in a given area.
We say that the SPEB of individual agents scales as $\Theta(f(n))$ for some function $f(n)$, denoted by $\PEB(\V{p}) \in \Theta(f(n))$, if there are deterministic constants $0<c_1<c_2<+\infty$ such that
    \begin{align}\label{eq:def_scaling}
        \mathbb{P} \left\{ c_1 f(n) \leq \PEB(\V{p})
        \leq c_2 f(n) \right\} = 1-\epsilon(n)\,, 
    \end{align}
where $\lim_{n\rightarrow \infty} \epsilon(n) = 0$.
\end{defi}

\begin{thm}\label{thm:Scaling_Law1}
In dense networks, the SPEB of each agent scales as $\Theta(1/\Nb)$
for non-cooperative localization, and as $\Theta(1/(\Nb+\Na))$ for
cooperative localization.
\end{thm}

\begin{IEEEproof}
See Appendix \ref{apd:scaling_laws}.
\end{IEEEproof}

\begin{thm}\label{thm:Scaling_Law2}
In extended networks with an amplitude loss exponent
$b$,\footnote{Note that the amplitude loss exponent is $b$, while
the corresponding power loss exponent is $2b$. The amplitude loss exponent $b$ is environment-dependent and can range from approximately 0.8 (e.g., hallways inside buildings) to 4 (e.g., dense urban environments) \cite{Gol:05}.}
the SPEB of each
agent scales as
    \begin{align*}
        \PEB(\V{p}) \in
            \begin{cases}
              \Theta(1/\log \Nb)\,, & b = 1,\\
              \Theta(1)\,,& b > 1,\\
              \Theta(1/N_\text{b}^{b-1})\,,& 0<b<1,
            \end{cases}
    \end{align*}
for non-cooperative localization, and
    \begin{align*}
        \PEB(\V{p}) \in
            \begin{cases}
              \Theta(1/\log (\Nb+\Na))\,, & b = 1,\\
              \Theta(1)\,,& b > 1,\\
              \Theta(1/(\Nb+\Na)^{b-1})\,,& 0<b<1,
            \end{cases}
    \end{align*}
for cooperative localization.
%
\end{thm}

\begin{IEEEproof}
See Appendix \ref{apd:scaling_laws}.
\end{IEEEproof}

\begin{rem} We make the following remarks.
\begin{itemize}
\item  In dense networks, the SPEB scales inversely
proportional to the number of anchors for non-cooperative
localization, and inversely proportional to the number of nodes for
cooperative localization. The gain from cooperation is given by
$\Theta(1+\Na/\Nb)$, and hence the benefit is most pronounced when
the number of anchors is limited. Moreover, it is proven in Appendix \ref{apd:scaling_laws} that $\epsilon(n)$ decreases exponentially with the number of nodes.

\item In extended networks with an amplitude loss exponent equal to 1, the SPEB scales inversely proportional to the logarithm of the
number of anchors for non-cooperative localization, and inversely
proportional to the logarithm of the number of nodes for cooperative
localization. This implies that the SPEB in extended networks
decreases much more slowly than that in dense networks, and
the gain from cooperation is now reduced to
$\Theta(\log(\Nb+\Na)/\log\Nb)$. Moreover, it is shown in Appendix \ref{apd:scaling_laws} that $\epsilon(n)$ decreases as $\exp(-(\log n)^2/8)/\log n$.

\item In extended networks with an amplitude loss exponent greater than 1, the SPEB converges to a strict positive value as the network grows. This agrees with our intuition that as more nodes are added, the benefit of the additional nodes diminishes due to the rapidly
decaying RII provided by those nodes. It can be shown that the SPEB  converges to a smaller value in the cooperative case than that in the non-cooperative case, i.e., a constant gain can be obtained by cooperation.
\end{itemize}
\end{rem}

\section{Numerical Results}\label{sec:Simu}

In this section, we examine several numerical examples pertaining to
cooperative localization and illustrate practical applications of
our analytical results.

\subsection{Effective Ranging Information}

\begin{figure}[t]
    \psfrag{AAAAAAA1}[l][][1.3]{\hspace{-7mm}$\phi_{1,2}=0$}
    \psfrag{AAAAAAA2}[l][][1.3]{\hspace{-7.5mm}$\phi_{1,2}=\pi/4$}
    \psfrag{AAAAAAA3}[l][][1.3]{\hspace{-7.5mm}$\phi_{1,2}=\pi/2$}
    \psfrag{AAAAAAA4}[l][][1.3]{\hspace{-7.5mm}Limits}

    \psfrag{xlabel}[c][][1.3]{RII $\nu_{1,2}$}
    \psfrag{ylabel}[c][][1.3]{Effective RII $\xi_{1,2} \, \nu_{1,2}$}
    \psfrag{title}[c][][1.3]{}

    \begin{center}
    {\includegraphics[angle=0,width=1\linewidth,draft=false]{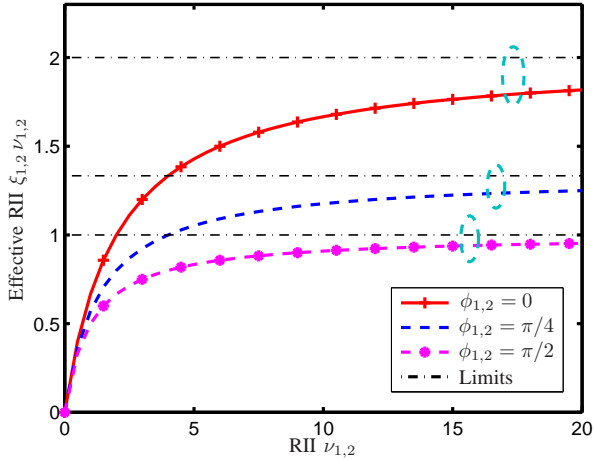}}

    \caption{\label{fig:Coop_new_2} Effective RII $\xi_{1,2} \, \nu_{1,2}$ as a function of the RII $\nu_{1,2}$, for $\V{J}_{\text{A},2} = \V{F}(\mu_2=2,\eta_2=1,\vartheta_2=0)$, and different angle of arrival $\phi_{1,2}$.
    }
    \end{center}
\end{figure}

We first investigate the behavior of the effective RII
$\tilde{\nu}_{1,2}$ from Corollary \ref{cor:SPEB_TwoAgent} when two
agents cooperate. The effective RII $\tilde\nu_{1,2}$ is plotted in
Fig.~\ref{fig:Coop_new_2} as a function of the RII $\nu_{1,2}$ for
$\JA{\V{p}_2}=\V{F}(\mu_2=2,\eta_2=1,\vartheta_2=0)$ and various
values of $\phi_{1,2}$. The corresponding asymptotic limits are also plotted for large values of $\nu_{1,2}$.
%
We observe that effective RII increases from 0 to $1/\Delta_2(\phi_{1,2})$ as the RII $\nu_{1,2}$ increases. For a fixed RII, the second agent will provide the maximum effective RII
at $\phi_{1,2} = \vartheta_2$, along which angle the second agent
has the minimum DPEB (i.e., $1/\mu_2 = 0.5$). On the other hand, the second agent will provide the minimum effective RII at $\phi_{1,2} = \vartheta_2 \pm \pi/2$, along which angle the second agent has the maximum DPEB (i.e., $1/\eta_2 = 1$).


\subsection{Benefit of Cooperation}\label{Sec:Sim_perf}

\begin{figure}[t]

    \psfrag{A}{}
    \psfrag{B}{}
    \psfrag{C}{}
    \psfrag{D}{}
    \psfrag{E}{}
    \psfrag{F}{}
    \psfrag{G}{}
    \psfrag{H}{}
    \psfrag{dist}[c][][1.8]{$D$}
    \psfrag{10}[c][][1.8]{$10$}
    \psfrag{-10}[c][][1.8]{$-10$}

    \begin{center}
{\includegraphics[angle=0,width=0.88\linewidth,draft=false]{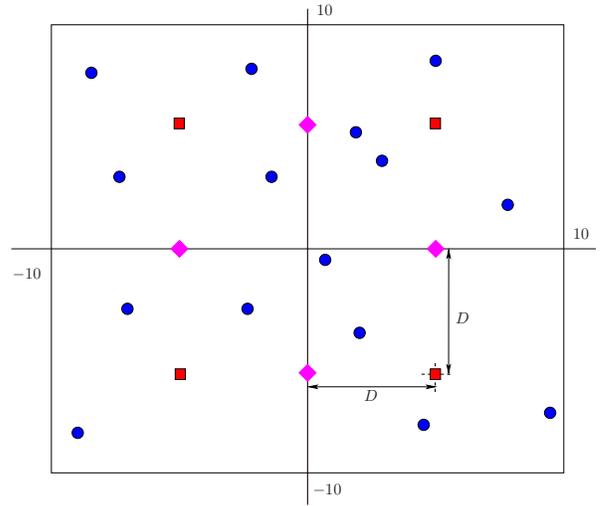}}

    \caption{\label{map_numerical} Typical network deployment of two sets
    of anchors (set I: squares, set II: diamonds) and $\Na=15$ agents.
    The agents are distributed uniformly over the $[-10,10\,] \times [-10,10\,]$
    map, while the locations of the anchors are controlled by $D$.}
    \end{center}
\end{figure}

We now consider the SPEB performance as a function of the number of agents for cooperative localization. The network configuration is shown in Fig.~\ref{map_numerical}. The agents randomly (uniformly distributed) reside in a 20\thinspace m by 20\thinspace m area. There are two sets of anchors (shown as squares (set I) and diamonds (set II) in Fig.~\ref{map_numerical}), with a configuration determined by the parameter $D$. Since fading does not affect the scaling behavior as shown Section \ref{sec:scaling_law}, we consider a network with signals that obey the free-space path-loss model for simplicity, so that the RII $\lambda_{kj} \propto 1/d_{kj}^2$.

\begin{figure}[t]
    \psfrag{AAAAAAA1}[l][][1.2]{\hspace{-9.5mm} 4 Anchors I}
    \psfrag{AAAAAAA2}[l][][1.2]{\hspace{-10mm} 4 Anchors II}
    \psfrag{AAAAAAA3}[l][][1.2]{\hspace{-10mm} 8 Anchors}
    \psfrag{AAAAAAA4}[l][][1.2]{\hspace{-10mm} Non-coop}

    \psfrag{xlabel}[c][][1.3]{Number of agents}
    \psfrag{ylabel}[c][][1.3]{$\mathbb{E} \left\{ \mathrm{SPEB} \right\} $ ($\rm{m}^2$)}
    \psfrag{title}[c][][1]{}

    \begin{center}
    {\includegraphics[angle=0,width=1\linewidth,draft=false]{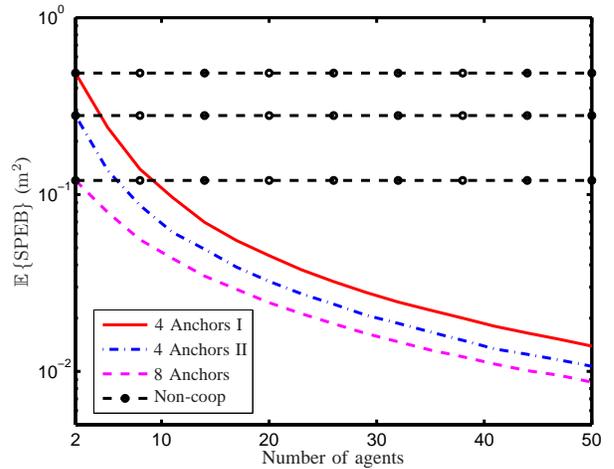}}
    \caption{\label{coop_bound} The average SPEB as a function of the
    number of agents in the network for various anchor configurations ($D=10$).}
    \end{center}
\end{figure}

Figure \ref{coop_bound} shows the average SPEB over all the agents as a function of the number of agents, obtained by Monte Carlo simulation, for $D=10$. We see that as the number of agents increases, the average SPEB decreases significantly, roughly proportional to the number of agents. Note that the anchor configuration set II yields a lower SPEB. Intuitively, this is due to the fact that the anchors in set II (distance $D$ from the center) cover the area better than the anchors in set I (distance $\sqrt{2}D$ from the center).

\begin{figure}[t]
    \psfrag{AAAAAAA1}[l][][1.2]{\hspace{-9.5mm} 4 Anchors I}
    \psfrag{AAAAAAA2}[l][][1.2]{\hspace{-10mm} 4 Anchors II}
    \psfrag{AAAAAAA3}[l][][1.2]{\hspace{-10mm} 8 Anchors}
    \psfrag{xlabel}[c][][1.3]{Number of agents}
    \psfrag{ylabel}[c][][1.3]{$\mathbb{E}\left\{ \PEB^\text{L}(\V{p})/\PEB^\text{U}(\V{p}) \right\}$}
    \psfrag{title}[c][][1.3]{}

    \begin{center}    {\includegraphics[angle=0,width=1\linewidth,draft=false]{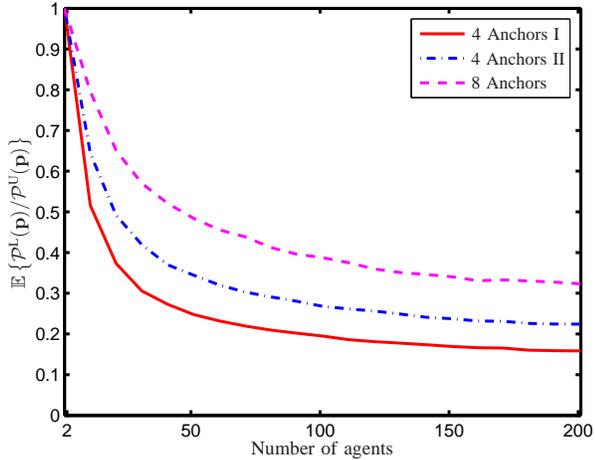}}

    \caption{\label{SPEB_appro_lin} Ratio of upper and lower approximations of the SPEB, $\PEB^\text{L}(\V{p})$ and $\PEB^\text{U}(\V{p})$, as a function of the number of agents for anchor set I, set II, and both.}
    \end{center}
\end{figure}

Define the upper and lower approximations of agent $k$'s SPEB as
    \begin{align*}
        \PEB^\text{U}(\V{p}_k) & \triangleq \T \left\{ \left[ \JL{\V{p}_k} \right]^{-1} \right\} \\ 
\noalign{ \noindent and \vspace{\jot}}
        \PEB^\text{L}(\V{p}_k) & \triangleq \T \left\{ \left[ \JU{\V{p}_k} \right]^{-1}\right\}\,,
    \end{align*}
where $\JL{\V{p}_k}$ and $\JU{\V{p}_k}$ are given by \eqref{eq:Anal_Lower} and \eqref{eq:Anal_Upper}, respectively, in Theorem \ref{pro:MultiAnt}. Figure \ref{SPEB_appro_lin} shows the average ratio of the lower and upper approximations of the SPEB, obtained by Monte Carlo simulation, for anchor set
I, set II, and both sets.
When there are only two agents in cooperation, the bounds coincide, as we expect from Corollary \ref{cor:SPEB_TwoAgent}.
As the number of agents increases, the ratio deviates from 1, or equivalently the approximations become looser, due to the fact that upper approximation ignores more cooperative information, and the lower approximation considers more agents to be equivalent to anchors. Nevertheless, the ratio converges to a positive constant, implying that the upper and lower approximation decrease at the same rate in an asymptotical regime, as shown in the proof of Theorem \ref{thm:Scaling_Law1}.

\subsection{Anchor Deployment}

\begin{figure}[t]
    \psfrag{AAAAAAA1}[l][][1.2]{\hspace{-9.5mm} 4 Anchors I}
    \psfrag{AAAAAAA2}[l][][1.2]{\hspace{-10mm} 4 Anchors II}
    \psfrag{AAAAAAA3}[l][][1.2]{\hspace{-10mm} 8 Anchors}
    \psfrag{AAAAAAA4}[l][][1.2]{\hspace{-10mm} Random (8)}

    \psfrag{xlabel}[c][][1.3]{$D$ (m)}
    \psfrag{ylabel}[c][][1.3]{$\mathbb{E} \left\{ \mathrm{SPEB} \right\} $ ($\rm{m}^2$)}
    \psfrag{title}[c][][1.3]{}
    \begin{center}
    {\includegraphics[angle=0,width=1\linewidth,draft=false]{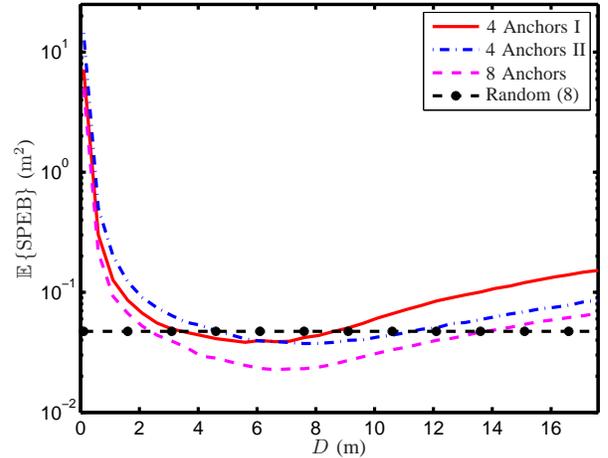}}

    \caption{\label{Bound_AncDeploy} The mean SPEB with respective to anchor deployment. There are $\Na=15$ agents.}
    \end{center}
\end{figure}

Finally, we investigate the effect of anchor deployment in more
detail. We consider a scenario with $\Na=15$ agents. The anchor
placement is controlled through $D$ (see Fig.~\ref{map_numerical}).
Figure~\ref{Bound_AncDeploy} shows the average SPEB as a function of
$D$ for different anchor configurations (set I, set II, and both sets). We see that the SPEB first decreases, and then increases,
as a function of $D$. When $D$ is close to 0, all the anchors are
located closely in the middle of the area, and hence the RIs
from those anchors to a particular agent are nearly in the same
direction. This will greatly increase the error of each agent's
position since every $\JA{\V{p}_k}$ is close to singular, resulting
in poor overall SPEB performance. As the anchors begin to move away
from the center, they provide RIs along different directions to each
agent, which lowers the average SPEB. Then, as the distances of the
anchors to the center increase further, the anchors become far away
from more and more agents. Hence the RII decreases due to the
path-loss phenomena, and this leads to the increase in the average
SPEB. Observe also that anchor set I is better than anchor set II
for $D<7 \thinspace \text{m}$. This is because, for a fixed $D<7 \thinspace \text{m}$, anchor set I can cover a larger area. For $D>7 \thinspace \text{m}$, anchor set I suffers more from path-loss than anchor set II.

For the sake of comparison, we have also included the average SPEB
when 8 anchors are deployed 1) according to set I and II simultaneously, and 2) randomly on a $[-10\thinspace \text{m},10\thinspace \text{m}\,] \times [-10\thinspace \text{m},10\thinspace \text{m}\,]$ area. The figure shows that
intelligent anchor deployment can be beneficial compared to
random deployment, indicating the need for anchor deployment
strategies.

\section{Conclusion}\label{sec:Conc}

In this paper, we have investigated the fundamental limits on the localization accuracy for wideband cooperative location-aware networks. We have derived the squared position error bound (SPEB) by applying the notion of equivalent Fisher information (EFI) to characterize the localization accuracy. Since our analysis exploits the received waveforms rather than specific signal metrics, the SPEB incorporates \emph{all} the localization information inherent in the received waveforms. Our methodology unifies the localization information from anchors and that from cooperation among agents in a canonical form, viz.\ ranging information (RI), and the total localization information is a sum of these individual RIs. We have put forth a geometrical interpretation of the EFIM based on eigen-decomposition, and this interpretation has facilitated the theoretical analysis of the localization information for cooperative networks.  We have also derived scaling laws for the SPEB in both dense and extended networks, showing the benefit of cooperation in an asymptotic regime.
Our results provide fundamental new insights into the essence of the
localization problem, and can be used as guidelines for localization
system design as well as benchmarks for cooperative location-aware
networks.


\appendices

\section{Proof of Proposition \ref{pro:SPEB_Decop}}
\label{apd:pro_SPEB_Decop}

\begin{IEEEproof}
The right-hand side of \eqref{eq:Anal_dirPEB} can be written as
    \begin{align*}
        \qquad \quad & \hspace{-4mm} \PEB(\V{p}_k;\V{u}) + \PEB(\V{p}_k;\V{u}_\bot) \\
        & = \T \left\{ \V{u}^\text{T}\,\left[ \JTH^{-1} \right]_{2 \times 2, k} \,\V{u} \right\}
        + \T \left\{ \V{u}_\bot^\text{T}\,\left[ \JTH^{-1} \right]_{2 \times 2, k} \,\V{u}_\bot
        \right\}\nonumber \\
        & = \T \left\{\left[ \JTH^{-1} \right]_{2 \times 2, k} \,\V{u}  \V{u}^\text{T}\right\}
        + \T \left\{\left[ \JTH^{-1} \right]_{2 \times 2, k}
        \,\V{u}_\bot \,\V{u}_\bot^\text{T}\right\} \nonumber\\
        & =\T \left\{\left[ \JTH^{-1} \right]_{2 \times 2, k}
        \right\} = \PEB(\V{p}_k)
        \, ,
    \end{align*}
where we have used the fact $\V{u} \,\V{u}^\text{T} +  \V{u}_\bot
\,\V{u}_\bot^\text{T} = \V{I}$.
\end{IEEEproof}

\begin{figure*}
	[!b] \vspace*{4pt} \hrulefill \normalsize \setcounter{MYtempeqncnt}{\value{equation}} \setcounter{equation}{28} 
	\begin{align}\label{eq:Kc_det}
	   \left[\, \V{K}_{\text{C}} \,\right]_{2k-1:2k,2m-1:2m}
	    & =   \begin{cases}
	            \sum_{j\in\NA\backslash\{ k\}} \left[\, \bm\Phi_{kj}
	            \left(\V{p}_{k},\V{p}_{k}\right) + \bm\Phi_{jk}
	            \left(\V{p}_{k},\V{p}_{k}\right)\, \right]
	            \, ,  &   k = m\,,\\
	            \bm\Phi_{km} \left(\V{p}_{k},\V{p}_{m}\right) + \bm\Phi_{mk}
	            \left(\V{p}_{k},\V{p}_{m}\right)
	            \, , &    k \neq m\,.
	        \end{cases}
	\end{align}
	\setcounter{equation}{\value{MYtempeqncnt}} \vspace*{-6pt} 
\end{figure*}

\begin{figure*}
	[!b] \vspace*{4pt} \hrulefill \normalsize \setcounter{MYtempeqncnt}{\value{equation}} \setcounter{equation}{29} 
	\begin{align}\label{eq:Mc_det}
	    \left[\, \V{M}_{\text{C}} \,\right]_{2k-1:2k,2m-1:2m}
	    & =   \begin{cases}
	            \sum_{j\in\NA\backslash\{ k\}} \left[\, \bm\Upsilon_{kj}
	            \left(\V{p}_{k},\bm{\kappa}_{kj},\V{p}_k \right) + \bm\Upsilon_{jk}
	            \left(\V{p}_{k},\bm{\kappa}_{jk},\V{p}_k \right)\,\right]
	            \, ,  &  k = m\,,\\
	            \bm\Upsilon_{km}\left(\V{p}_{k},\bm{\kappa}_{km}, \V{p}_m\right) + \bm\Upsilon_{mk}\left(\V{p}_{k},\bm{\kappa}_{mk}, \V{p}_m \right)
	            \, , &  k \neq m\,.
	        \end{cases}
	\end{align}
	\setcounter{equation}{\value{MYtempeqncnt}} \vspace*{-6pt} 
\end{figure*}

\section{Proof of Theorem \ref{thm:EFIM_Coop}}\label{apd:EFIM_Coop}

We proceed in two steps: we first show that the EFIM is structured
as in \eqref{eq:Coop_EFIM}, and then derive the details of the RI.

\subsection{Derivation of the EFIM Structure} \label{apd:EFIM_Coop_str}

When {a priori} knowledge of the agents' positions is unavailable,
the log-likelihood function in \eqref{eq:Eval_LN} becomes
    \begin{align}\label{eq:Apd_LRT_1}
        \ln f(\V{r},\bm{\kappa}|\V{P})
        = \sum_{k \in \NA} \; \sum_{j \in \NB\cup\NA\backslash\{k\}}
        {\Big[}\ln & f(\V{r}_{kj}|\V{p}_k, \V{p}_j, \bm\kappa_{kj}) \nonumber \\
& \hspace{-10mm} + \ln f(\bm\kappa_{kj} | \V{p}_k, \V{p}_j) {\Big]}
        \, ,
    \end{align}
where $\bm{\kappa}$ denotes the vector of the channel parameters containing all $\bm{\kappa}_{kj}$ with $k \in \NA$ and $j \in
\NB\cup\NA\backslash\{k\}$. For notational convenience, we now
introduce
    \begin{align}\label{eq:apd_def_phi}
        \bm\Phi(\V{x},\V{y}) & \triangleq \E_{\V{r},\bm{\kappa}} \left\{ - \frac{\partial^2 {\ln f(\V{r},\bm{\kappa}|\V{P})}} {\partial \V{x} \partial \V{y}^\text{T}}
        \right\}\, ,\\
        \bm\Phi_{kj}(\V{x},\V{y}) & \triangleq
        \E_{\V{r},\bm{\kappa}} {\bigg\{} - \frac{\partial^2}{\partial \V{x} \partial \V{y}^\text{T}}
        {\Big[} \ln f(\V{r}_{kj}|\V{p}_k, \V{p}_j, \bm\kappa_{kj}) \nonumber \\
        & \hspace{30mm} + \ln f(\bm\kappa_{kj}|\V{p}_k, \V{p}_j) {\Big]} {\bigg\}} \, ,\label{eq:apd_def_phi_xy}
    \end{align}
as well as
    \begin{align*}
        \bm{\Upsilon}(\V{x},\V{y},\V{z}) & \triangleq \bm\Phi(\V{x},\V{y})
        \left[\bm\Phi(\V{y},\V{y})\right]^{-1}
        \bm\Phi(\V{y},\V{z})\, , \\
        \bm{\Upsilon}_{kj}(\V{x},\V{y},\V{z}) & \triangleq \bm\Phi_{kj}(\V{x},\V{y})
        \left[\bm\Phi_{kj}(\V{y},\V{y})\right]^{-1}
        \bm\Phi_{kj}(\V{y},\V{z})
        \,.
    \end{align*}
Since $\bm\Phi(\tilde{\bm\theta}_k,\tilde{\bm\theta}_j) = \V{0}$ for
$k \neq j$, the EFIM for $\V{P}$ can be derived as
    \begin{align}\label{eq:JE_EFIM_Deriv}
        \JE{\V{P}} = \bm\Phi\left( \V{P}, \V{P} \right) - \sum_{k \in \NA}
        \bm{\Upsilon}\left( \V{P}, \tilde{\bm\theta}_k, \V{P}\right)
        \, .
    \end{align}

\subsubsection*{Structure of $\bm\Phi\left(\V{P},\V{P}\right)$}

Due to the structure in \eqref{eq:Apd_LRT_1}, we can express
$\bm\Phi\left(\V{P},\V{P}\right)$ as
    \begin{align*}
        \bm\Phi\left(\V{P},\V{P}\right)
        & = \sum_{k \in \NA} \sum_{j\in \NB } \bm\Phi_{kj}\left(\V{P},\V{P}\right)
        + \sum_{k \in \NA} \sum_{j\in \NA\backslash\{k\} } \bm\Phi_{kj}\left(\V{P},\V{P}\right)
        \nonumber \\
        & \triangleq \V{K}_\text{A} + \V{K}_\text{C}
        \, ,
    \end{align*}
where $\V{K}_{\text{A}} \in \mathbb{R}^{2\Na \times 2\Na}$ is a
block-diagonal matrix, consisting of $2\times2$ block matrices,
given by
    \begin{align*}
        \left[\, \V{K}_{\text{A}}\, \right]_{2k-1:2k,2m-1:2m}
        =   \begin{cases}
                \sum_{j\in\NB}\B\Phi_{kj}\left(\V{p}_{k},\V{p}_{k}\right)
                \, ,  &   k = m\,,\\
                \V{0}\, , &   k \neq m\,.
            \end{cases}
    \end{align*}
On the other hand, $\V{K}_{\text{C}} \in \mathbb{R}^{2\Na \times
2\Na}$ is also a block-matrix, consisting of $2\times2$ block
matrices, given by (\ref{eq:Kc_det}) shown at the bottom of the page. \addtocounter{equation}{1}



\subsubsection*{Structure of $\bm{\Upsilon}\left( \V{P}, \tilde{\bm\theta}_k, \V{P} \right)$}

Since $\bm{\Phi}\left(\bm\kappa_{ki},\bm\kappa_{kj}\right) = \V{0}$
for $i \neq j$, we find that
    \begin{align*}
        \sum_{k\in\NA} \bm{\Upsilon}\left(\V{P}, \tilde{\bm\theta}_k, \V{P} \right)
        & = \sum_{k \in \NA} \sum_{j\in \NB }
        \bm{\Upsilon}_{kj} \left(\V{P},\bm\kappa_{kj},\V{P} \right) \\
        & \hspace{5mm} + \sum_{k \in \NA} \sum_{j\in \NA\backslash\{k\} }
        \bm{\Upsilon}_{kj} \left(\V{P},\bm\kappa_{kj},\V{P} \right)
        \nonumber \\
        & \triangleq \V{M}_\text{A} + \V{M}_\text{C} \, ,
    \end{align*}
where $\V{M}_\text{A} \in \mathbb{R}^{2\Na \times 2\Na}$ is a
block-diagonal matrix, consisting of $2\times2$ block matrices,
given by
    \begin{align*}
        & \hspace{-4mm} \left[\, \V{M}_{\text{A}}\, \right]_{2k-1:2k,2m-1:2m} \\
        & =   \begin{cases}           \sum_{j\in\NB}\bm\Upsilon_{kj}\left(\V{p}_{k},\bm\kappa_{kj},
                \V{p}_k\right)\, ,  &   k = m\,,\\
                \V{0}\, ,           &  k \neq m\,.
            \end{cases}
    \end{align*}
On the other hand, $\V{M}_\text{C} \in \mathbb{R}^{2\Na \times 2\Na}
$ is also a block-matrix, consisting of $2\times2$ block matrices,
given by (\ref{eq:Mc_det}) shown at the bottom of the page. \addtocounter{equation}{1}


\subsubsection*{Structure of $\V{J}_\text{\emph{e}}(\V{P})$}

Combining these results, we find that the EFIM in
\eqref{eq:JE_EFIM_Deriv} can be written as
    \begin{align}\label{eq:Apd_JE_1}
        \JE{\V{P}} & = \underbrace{{\Big\{} \V{K}_\text{A} - \V{M}_\text{A} {\Big\}}}_\text{from anchors}
        \; + \underbrace{{\Big\{} \V{K}_\text{C} -  \V{M}_\text{C} {\Big\}}}_\text{from cooperation}
        \, ,
    \end{align}
from which we obtain \eqref{eq:Coop_EFIM}. In
\eqref{eq:Coop_EFIM}, $\JA{\V{p}_k} = \sum_{j\in \NB}
\V{R}_k(\V{r}_{kj})$ and $\C_{kj} = \C_{jk} = \V{R}_k(\V{r}_{kj})
+ \V{R}_k(\V{r}_{jk})$ in which we have introduced the RI:
    \begin{align}\label{eq:Apd_RI}
        \V{R}_k(\V{r}_{kj}) = \bm\Phi_{kj}\left( \V{p}_k, \V{p}_k\right)
        - \bm\Upsilon_{kj} \left(\V{p}_{k},\bm{\kappa}_{kj},\V{p}_k \right)
        \, .
    \end{align}
Note that in the derivation, we used $$\bm\Phi_{km}
\left(\V{p}_{k},\V{p}_{m}\right) = - \bm\Phi_{km}
\left(\V{p}_{k},\V{p}_{k}\right)\,,$$ and
$$\bm\Upsilon_{km}\left(\V{p}_{k},\bm{\kappa}_{km}, \V{p}_m \right) =
-\bm\Upsilon_{km}\left(\V{p}_{k},\bm{\kappa}_{km}, \V{p}_k \right)\,.$$

Since $\JE{\V{P}}$ in \eqref{eq:Coop_EFIM} can be expressed in terms of the RIs $\V{R}_k(\V{r}_{kj})$, for $k \in \NA$ and $j \in \NB\cup\NA\backslash\{k\}$, we will examine next the details of the RIs.

\begin{figure*}
	[!b] \vspace*{0pt} \hrulefill \normalsize \setcounter{MYtempeqncnt}{\value{equation}} \setcounter{equation}{34} 
    \begin{align}\label{eq:Apd_RI_Inten}
        \lambda_{kj} & \triangleq  \frac{1}{c^2} {\Bigg[} \, \V{l}_{kj}^\text{T} \, \bm\Psi_{kj} \, \V{l}_{kj}
        + c^2 \Xi_{kj}(d_{kj},d_{kj})  \nonumber \\
        & \hspace{10mm}
        - {\Big(} \V{l}_{kj}^\text{T} \, \bm\Psi_{kj} + c^2 \bm\Xi_{kj}(d_{kj},\bm\kappa_{kj}) {\Big)}
            {\Big(} \bm\Psi_{kj}+ c^2 \bm\Xi_{kj}(\bm\kappa_{kj},\bm\kappa_{kj}) {\Big)}^{-1}
            {\Big(} \V{l}_{kj}^\text{T} \, \bm\Psi_{kj} + c^2 \bm\Xi_{kj}(d_{kj},\bm\kappa_{kj}) {\Big)}^\text{T}
        {\Bigg]}
    \end{align}
	\setcounter{equation}{\value{MYtempeqncnt}} \vspace*{-6pt} 
\end{figure*}

\begin{figure*}
	[!b] \vspace*{0pt} \hrulefill \normalsize \setcounter{MYtempeqncnt}{\value{equation}} \setcounter{equation}{35} 
	\begin{align}\label{eq:Apd_RII_Limit}
	    \lambda_{kj} = \frac{1}{c^2} \; \V{l}_{kj}^\text{T} \, \bm\Psi_{kj} \, {\Big(} \bm\Psi_{kj} + \bm\Xi_{kj}(\bm\kappa_{kj},\bm\kappa_{kj}){\Big) }^{-1} \, \bm\Xi_{kj}(\bm\kappa_{kj},\bm\kappa_{kj}) \, \V{l}_{kj}
	\end{align}
	\setcounter{equation}{\value{MYtempeqncnt}} \vspace*{-6pt} 
\end{figure*}

\subsection{Details of the Ranging Information} \label{apd:EFIM_Coop_RII}

We now consider the detailed expression of the RI
$\V{R}_{k}\left(\V{r}_{kj}\right)$ in \eqref{eq:Apd_RI}. We first
introduce
    \begin{align}
        \bm\Xi_{kj}(\V{x},\V{y}) & \triangleq
            \E_{\bm{\kappa}} \left\{ - \frac{\partial^2 \ln
            f(\bm\kappa_{kj}|\V{p}_k,\V{p}_j)}{\partial \V{x} \,
            \partial \V{y}^\text{T}} \right\} \, ,\nonumber \\
        \noalign{\noindent and \vspace{\jot}}
        \bm\Psi_{kj} & \triangleq \E_{\V{r},\bm{\kappa}} \left\{ -
            \frac{\partial^2
            {\ln f(\V{r}_{kj}|\V{p}_k,\V{p}_j,\bm\kappa_{kj})}} {\partial \tilde{\bm\kappa}_{kj} \,
            \partial \tilde{\bm\kappa}_{kj}^\text{T}} \right\}\, , \label{eq:Def_Psi_kj}
    \end{align}
where $\tilde{\bm\kappa}_{kj} = {\big[} \,\D{kj}{1} \;\;
\tilde{\alpha}_{kj}^{(1)} \;\; \D{kj}{2} \;\;
\tilde{\alpha}_{kj}^{(2)} \cdots  \;\; \D{kj}{L_{kj}} \;\;
\tilde{\alpha}_{kj}^{(L_{kj})} \,{\big]}^\text{T}$ with
$\tilde{\alpha}_{kj}^{(l)} \triangleq \A{kj}{l}/c$.

From \eqref{eq:Model_Tau_L} and \eqref{eq:Anal_PDFkj}, we note that
$d_{kj}=\left\Vert \V{p}_{k}-\V{p}_{j}\right\Vert$ and that $f\left(\V{r}_{kj}|\V{p}_{k},\V{p}_{j},
\bm{\kappa}_{kj}\right)$ and
$f\left(\bm{\kappa}_{kj}|\V{p}_{k},\V{p}_{j}\right)$ only depend on
$\V{p}_{k},\V{p}_{j}$ through $d_{kj}$. Using the chain rule, we
have
    \begin{align*}
        \bm\Phi_{kj}\left( \V{p}_k, \V{p}_k\right) & =
        \frac{\partial d_{kj}}{\partial \V{p}_{k}} \, \Phi_{kj}\left(d_{kj},d_{kj}\right)
        \, \frac{\partial d_{kj}}{\partial \V{p}_{k}^\text{T}} \, , \\
        \noalign{\noindent and \vspace{\jot}}
        \bm\Upsilon_{kj} \left(\V{p}_{k},\bm{\kappa}_{kj},\V{p}_k
        \right) & = \frac{\partial d_{kj}}{\partial \V{p}_{k}} \, \Upsilon_{kj}\left(d_{kj},\bm{\kappa}_{kj},d_{kj}\right)
        \, \frac{\partial d_{kj}}{\partial \V{p}_{k}^\text{T}}\, ,
    \end{align*}
and hence $\V{R}_{k}\left(\V{r}_{kj}\right)$ can be expressed as
    \begin{align}\label{eq:apd_RII}
        \V{R}_{k}\left(\V{r}_{kj}\right)
        & =  \Phi_{kj}\left(d_{kj},d_{kj}\right) \, \V{q}_{kj}\, \V{q}_{kj}^\text{T} \nonumber \\
        & \hspace{4mm} - \Upsilon_{kj}\left(d_{kj},\bm{\kappa}_{kj},d_{kj}\right) \, \V{q}_{kj}\,\V{q}_{kj}^\text{T}
        \nonumber \\
        & = \lambda_{kj} \, \V{q}_{kj}\, \V{q}_{kj}^\text{T}
        \, ,
    \end{align}
where $\V{q}_{kj} \triangleq \partial d_{kj}/\partial\V{p}_{k} = -
\partial d_{kj}/\partial\V{p}_{j} = \left[\, \cos\phi_{kj} \;\;
\sin\phi_{kj} \, \right]^\text{T}$, 
and $\lambda_{kj}$ is given by (\ref{eq:Apd_RI_Inten}) shown at the bottom of the page, \addtocounter{equation}{1}
where $\V{l}_{kj} \triangleq {\underbrace{\Matrix{ccccc} {1  & 0 &
\cdots &  1 & 0}}_{2L_{kj}}}^\text{T} $.

\begin{figure*}
	[!b] \vspace*{4pt} \hrulefill \normalsize \setcounter{MYtempeqncnt}{\value{equation}} \setcounter{equation}{38} 
	\begin{align}\label{eq:limit_rii}
	    \lim_{t^2\rightarrow \infty} \left(
	    \sum_{j\in\NB} \V{R}_{\Na}(\V{r}_{\Na,j}) +
	    \sum_{j\in\NA\backslash\{\Na\}}
	    \left[\, \V{R}_{\Na}(\V{r}_{\Na,j})+
	    \V{R}_{\Na}(\V{r}_{j,\Na}) \,\right] +
	    \Matrix{cc}{t^2 \\ & t^2} \right)^{-1} = \V{0}
	\end{align}
	\setcounter{equation}{\value{MYtempeqncnt}} \vspace*{-6pt} 
\end{figure*}

\begin{figure*}
	[!b] \vspace*{4pt} \hrulefill \normalsize \setcounter{MYtempeqncnt}{\value{equation}} \setcounter{equation}{41} 
	\begin{align}\label{eq:Anal_EFIM_Coop_Lower}
	    \JL{\V{P}} = \Matrix{cccc}
	        {   \JA{\V{p}_1} + \sum\limits_{j\in\NA\backslash\{1\}}\C_{1,j}     &   -\C_{1,2}              &   \cdots  &   -\C_{1,\Na}  \\
	                -\C_{1,2}                                       &   \JA{\V{p}_2} + \C_{1,2}     &         &  {\Large \text{0}}  \\
	                \vdots      &      &   \ddots  &       \\
	                -\C_{1,\Na}    &    {\Large \text{0}}    &    &  \JA{\V{p}_\Na} + \C_{1,\Na} } 
	\end{align}
	\setcounter{equation}{\value{MYtempeqncnt}} \vspace*{-6pt} 
\end{figure*}

\section{Proof of Theorem \ref{cor:RII_NoPrior}} \label{apd:EFIM_NoPrior}

\begin{IEEEproof}
When {a priori} channel knowledge is unavailable, we have
$\Xi_{kj}(d_{kj},d_{kj})=0$, and $\bm\Xi_{kj}(d_{kj},\bm\kappa_{kj})
= \V{0}$. For NLOS signals, the RII in \eqref{eq:Apd_RI_Inten}
becomes $\lambda_{kj}=0$ since
$\bm\Xi_{kj}(\bm\kappa_{kj},\bm\kappa_{kj}) = \V{0}$. For LOS
signals, however, after some algebra, the RII becomes (\ref{eq:Apd_RII_Limit}) shown at the bottom of the page, \addtocounter{equation}{1}
where $\bm\Xi_{kj}(\bm\kappa_{kj},\bm\kappa_{kj}) = \lim_{t^2
\rightarrow \infty}\Diag{t^2, \V{0}}$ since the Fisher information
for known $\BB{kj}{1}=0$ is infinity. To simplify
\eqref{eq:Apd_RII_Limit}, we partition $\bm\Psi_{kj}$ as
    \begin{align*}
        {\bm\Psi}_{kj} =
            \Matrix{cc} {   u_{kj}^2   &   \V{k}_{kj}^\text{T} \\
                            \V{k}_{kj} &   {\breve{\B\Psi}}_{kj}} ,
    \end{align*}
where $u_{kj}^2 = 8\pi^2\beta^2 \, \mathsf{SNR}_{kj}^{(1)}$
obtained from \eqref{eq:Def_Psi_kj} through some algebra. As $t^2
\rightarrow \infty$ in \eqref{eq:Apd_RII_Limit}, we have
    \begin{align*}
        \lambda_{kj} & = \frac{8\pi^2\beta^2}{c^2} \, (1 - \chi_{kj})\, \mathsf{SNR}_{kj}^{(1)} \, ,
    \end{align*}
where
    \begin{align}\label{eq:Apd_chi_a}
        \chi_{kj} \triangleq \frac{{\V{k}_{kj}^\text{T}\,{\breve{\B\Psi}}_{kj}^{-1}\,
        \V{k}_{kj}}}{ {8\pi^2\beta^2 \, \mathsf{SNR}_{kj}^{(1)}} }
    \end{align}
is called {path-overlap coefficient} \cite{SheWin:J10a}.

We next show that only the first contiguous-cluster contains
information for localization. Let us focus on $\chi_{kj}$. If the
length of the first contiguous-cluster in the received waveform is
$\tilde{L}_{kj}$, where $1\leq\tilde{L}_{kj}\leq L_{kj}$, we have \cite{SheWin:J10a}
    \begin{align*}
        \V{k}_{kj} \triangleq \Matrix{cc} { \tilde{\V{k}}_{kj}^\text{T}   &
        \B{0}^\text{T}}^\text{T} \quad \text{and} \quad
        {\breve{\B\Psi}}_{kj} \triangleq
        \Matrix{cc} {   {\tilde{\B\Psi}}_{kj}    &   \B{0}   \\
                        \B{0}                   &   \boxtimes } ,
    \end{align*}
where  $\tilde{\V{k}}_{kj} \in \mathbb{R}^{2\tilde{L}_{kj}-1}$,
$\tilde{\B\Psi}_{kj} \in \mathbb{R}^{(2\tilde{L}_{kj}-1) \times
(2\tilde{L}_{kj}-1)}$, and $\boxtimes$ is a block matrix that is irrelevant
to the rest of the derivation. Hence \eqref{eq:Apd_chi_a} becomes
    \begin{align*}
        \chi_{kj} =  \frac{\tilde{\V{k}}_{kj}^\text{T} \:
        \tilde{\B\Psi}_{kj}^{-1} \: \tilde{\V{k}}_{kj}}{{8\pi^2\beta^2 \, \mathsf{SNR}_{kj}^{(1)}}}\: 
        \, ,
    \end{align*}
which depends only on the first $\tilde{L}_{kj}$ paths, implying that only the first contiguous-cluster of LOS signals contains information for localization.
\end{IEEEproof}



\section{Proof of Theorem \ref{thm:EFIM_Prior} and Corollary
\ref{thm:EFIM_Coop_Prior}}\label{sec:Apd_coop_Prior}

\begin{IEEEproof}
When the a priori knowledge of the agents' position is available,
the derivation of EFIM, equation \eqref{eq:Apd_LRT_1} becomes
    \begin{align*}
        \ln\LrT & = \sum_{k \in \NA} \: \sum_{j \in \NB\cup\NA\backslash\{k\}}
        {\bigg[} \ln f(\V{r}_{kj}|\V{p}_k, \V{p}_j, \bm\kappa_{kj}) \\
& \hspace{25mm} + \ln f(\bm\kappa_{kj} | \V{p}_k, \V{p}_j) {\bigg]}
        + \ln f\left(\V{P}\right) .
    \end{align*}
Following the notations and derivations in Appendix
\ref{apd:EFIM_Coop_str}, we obtain the EFIM given by
\eqref{eq:apd_EFIM_Prior}. This completes the proof of Theorem
\ref{thm:EFIM_Prior}. Note that the structure of
\eqref{eq:apd_EFIM_Prior} is similar to that of \eqref{eq:Apd_JE_1}
except the additional term $\bm\Xi_\V{P}$.

The EFIM in \eqref{eq:apd_EFIM_Prior} is applicable to general case.
Note that $\V{R}_k(\V{r}_{kj})$ in this case cannot be further
simplified as that in \eqref{eq:apd_RII} since we need to take
expectation over the random parameter $\V{P}$ in \eqref{eq:Apd_RI}.
However, when condition \eqref{eq:conditionPrior} holds for
functions $\Phi_{kj}\left(d_{kj},d_{kj}\right) \, \V{q}_{kj}\,
\V{q}_{kj}^\text{T}$, $\V{q}_{kj}\, \bm\Phi_{kj}(d_{kj},\V{p}_k)$, and
$\bm\Phi_{kj}(\bm{\kappa}_{kj},\bm{\kappa}_{kj})$, the expectations
of those functions with respect to $\V{P}$ can be replaced by the
values of the functions at $\bar{\V{P}}$. In such a case, the RI in
\eqref{eq:apd_RII_1} can be written as
    \begin{align*}
        \V{R}_k(\V{r}_{kj}) = \bar\lambda_{kj}\:\R(\bar\phi_{kj})
        \, ,
    \end{align*}
where $\bar\lambda_{kj}$ is the RII given in \eqref{eq:Apd_RI_Inten}
evaluated at $\bar{\V{P}}$, and $\bar\phi_{kj}$ is the angle from
$\bar{\V{p}}_k$ to $\bar{\V{p}}_j$.
\end{IEEEproof}


\begin{figure*}
	[!b] \vspace*{0pt} \hrulefill \normalsize \setcounter{MYtempeqncnt}{\value{equation}} \setcounter{equation}{44} 
	\begin{align}\label{eq:Anal_EFIM_Coop_Upper}
	    \hspace{-3mm}
	    \JU{\V{P}} = \Matrix{cccc}
	        {  \JA{\V{p}_1} + \!\!\!\! \sum\limits_{j\in\NA\backslash\{1\}}\C_{1,j}     &   -\C_{1,2}              &   \cdots  &   -\C_{1,\Na}  \\
	            -\C_{1,2}          &   \JA{\V{p}_2} + \C_{1,2}  + \!\!\!\! \sum\limits_{j\in\NA\backslash\{1,2\}}2\,\C_{2,j}     &      &   {\Large \text{0}}  \\
	            \vdots           &        &   \ddots  &        \\
	            -\C_{1,\Na}      &    {\Large \text{0}}    &    &  \JA{\V{p}_\Na} + \C_{1,\Na}  + \!\!\!\! \sum\limits_{j\in\NA\backslash\{1,\Na\}}2\,\C_{\Na,j} } 
	\end{align}
	\setcounter{equation}{\value{MYtempeqncnt}} \vspace*{-6pt} 
\end{figure*}

\section{Proof of Theorem \ref{cor:anc_agent}}\label{sec:Apd_anc_agent}

\begin{IEEEproof}
Consider a cooperative network with $\Na$ agents, whose overall EFIM
is given by \eqref{eq:apd_EFIM_Prior}. If agent $\Na$ has infinite
{a priori} position knowledge, i.e., $\B\Xi_{\V{p}_\Na} = \lim_{t^2
\rightarrow \infty} \Diag{t^2,t^2} $, then we apply the notion of
EFI to eliminate the parameter vector $\V{p}_\Na$ in
\eqref{eq:apd_EFIM_Prior} and have
    \begin{align}\label{eq:Anchor_Agent}
        \JE{\V{p}_1,\ldots,\V{p}_{\Na-1}} =
        \left[\, \JE{\V{P}}\, \right]_{2(\Na-1) \times 2(\Na-1)}
        \, ,
    \end{align}
where we have used (\ref{eq:limit_rii}) shown at the bottom of the page. \addtocounter{equation}{1}

Note that if we let $\mathcal{N}_\text{b}' \triangleq \NB \cup \{\Na\}$, $\mathcal{N}_\text{a}'
\triangleq \NA \backslash \{\Na\}$, and $\V{R}_k'(\V{r}_{k,\Na})
= \V{R}_{k}(\V{r}_{\Na,k}) + \V{R}_{k}(\V{r}_{k,\Na})$ for
$k\in\NA'$ in \eqref{eq:Anchor_Agent}, the structure of
\eqref{eq:Anchor_Agent} becomes the same as that of
\eqref{eq:apd_EFIM_Prior}, with a dimension decrease by 2.
Therefore, the new RI $\V{R}_k'(\V{r}_{k,\Na})$ is fully
utilizable, i.e., agent $\Na$ with infinite a priori position
knowledge is effectively an anchor.
\end{IEEEproof}

\section{Proofs for Section \ref{Sec:Geom}}
\label{Sec:Apd_pro}

\subsection{Proof of Proposition \ref{pro:SPEB_CIndep}}

\begin{IEEEproof}
If the current coordinate system is rotated by angle $\phi$ and
translated by $\V{p}_0 = [\, x_0 \; y_0 \,]^\text{T}$, then the
position of the agent in the new coordinate system is $\tilde{\V{p}}
= \U_\phi \,\V{p} + \V{p}_0$.
Consequently, the EFIM for $\tilde{\V{p}}$ is
    \begin{align}\label{eq:Anal_Coor_Sub}
        \JE{\tilde{\V{p}}}  &=
        \left[ \frac{\partial \V{p}}{\partial \tilde{\V{p}}} \right]^\text{T}
        \,   \JE{{\V{p}}} \,
        \left[ \frac{\partial \V{p}}{\partial \tilde{\V{p}}} \right]
        \nonumber \\
        &=  \U_\phi^\text{T} \, \JE{{\V{p}}} \, \U_\phi
        \,.
    \end{align}
Due to the cyclic property of the trace operator \cite{HorJoh:B85},
we immediately find that
    \begin{align}
        \PEB(\tilde{\V{p}})=\T \left\{ \left[\JE{\tilde{\V{p}}}\right]^{-1}
            \right\}= \T \left\{ \left[\JE{{\V{p}}}\right]^{-1}
            \right\}=\PEB(\V{p})
        \, .
    \end{align}
\end{IEEEproof}

\subsection{Proof of Proposition \ref{pro:MultiAnt}} \label{Sec:Apd_multiAnt}

\begin{IEEEproof}
Without loss of generality, we focus on the first agent.

\subsubsection*{Lower Bound}
Consider the EFIM $\JL{\V{P}}$ shown in (\ref{eq:Anal_EFIM_Coop_Lower}) at the bottom of the page.\addtocounter{equation}{1} It can be obtained from $\JE{\V{P}}$ by setting all $\C_{kj}=\V{0}$ for $1< k,\,j \leq \Na$. This EFIM corresponds to the situation where cooperation among  agents 2 to $\Na$ is completely ignored. 
One can show using elementary algebra that $\JL{\V{P}} \preceq {\V{J}}_\text{e}(\V{P})$, which agrees with intuition since the cooperation information among agents 2 to $\Na$ is not exploited. Applying the notion of EFI, we have the EFIM for the first agent as
    \begin{align*}
        \JL{\V{p}_1} & = \JA{\V{p}_1} \\ 
& \;\;\; + \!\! \sum_{j\in\NA\backslash\{1\}}
        {\Big[} \C_{1,j} - \C_{1,j} \, \left( \JA{\V{p}_j} + \C_{1,j}\right)^{-1} \, \C_{1,j} {\Big]} .
    \end{align*}
Since $\C_{1,j} = \nu_{1,j} \, \q_{\phi_{1,j}}
\q_{\phi_{1,j}}^\text{T}$ where $\q_{\phi_{1,j}} \triangleq [\,
\cos\phi_{1,j} \;\; \sin\phi_{1,j} \, ]^\text{T}$, we can express
$\JL{\V{p}_1}$ as
    \begin{align}\label{eq:Apd_Lapp_agent1}
        \JL{\V{p}_1} & = \JA{\V{p}_1} + \sum_{j\in\NA\backslash\{1\}} \xi_{1,j}^{\text{L}} \, \C_{1,j}
        \, ,
    \end{align}
where $\xi_{1,j}^{\text{L}}  \triangleq 1 - \nu_{1,j} \,
\q_{\phi_{1,j}}^\text{T} \left( \JA{\V{p}_j} + \C_{1,j} \right)^{-1}
\q_{\phi_{1,j}}$. The coefficient $\xi_{1,j}^{\text{L}}$ can be
simplified as
    \begin{align}\label{eq:apd_chi_L}
        \xi_{1,j}^{\text{L}} &  = 1 - \nu_{1,j} \, \q_{\vartheta_j-\phi_{1,j}}^\text{T} \nonumber \\ 
& \quad\; \cdot {\Bigg(} \Matrix{cc}{\mu_{j} &\\& \eta_j}
        + \nu_{1,j} \, \q_{\vartheta_j-\phi_{1,j}} \q_{\vartheta_j-\phi_{1,j}}^\text{T} {\Bigg)}^{-1} \!\! \q_{\vartheta_j - \phi_{1,j}}
        \nonumber \\
        & = \frac{1}{1 + \nu_{1,j} \, \Delta_j (\phi_{1,j})}
        \, ,
    \end{align}
where
    \begin{align*}
        {\Delta}_j(\phi_{1,j}) &= \frac{1}{{\mu}_j} \, \cos^2\left(\vartheta_j-\phi_{1,j} \right)
        + \frac{1}{{\eta}_j} \, \sin^2\left(\vartheta_j-\phi_{1,j}\right)
        \, .
    \end{align*}

\subsubsection*{Upper Bound}
Consider the EFIM $\JU{\V{P}}$ shown in (\ref{eq:Anal_EFIM_Coop_Upper}) at the bottom of the next page. It can be obtained from $\JE{\V{P}}$ by doubling the diagonal elements $\C_{kj}$ and setting the off-diagonal elements $-\C_{kj} = \V{0}$ for $1< k,\,j \leq \Na$. \addtocounter{equation}{1}
One can show using elementary algebra that $\JU{\V{P}} \succeq {\V{J}}_\text{e}(\V{P})$, which agrees with intuition since more cooperation information among agents 2 to $\Na$ is assumed in \eqref{eq:Anal_EFIM_Coop_Upper}. Applying the notion of EFI and following the similar analysis leading to \eqref{eq:Apd_Lapp_agent1} and \eqref{eq:apd_chi_L}, we obtain the EFIM for agent 1 as
    \begin{align*}
        \JU{\V{p}_1} = \JA{\V{p}_1} + \sum_{j\in\NA\backslash\{1\}} \xi_{1,j}^\text{U} \, \C_{1,j} \, ,
    \end{align*}
where
    \begin{align}\label{eq:apd_chi_U}
        \xi_{1,j}^\text{U} = \frac{1}{1 + \nu_{1,j} \,
        \tilde{\Delta}_j(\phi_{1,j})} \,,
    \end{align}
in which
    \begin{align*}
        \tilde{\Delta}_j(\phi_{1,j}) & = \frac{1}{\tilde{\mu}_j} \, \cos^2\left(\tilde\vartheta_j-\phi_{1,j} \right)
        + \frac{1}{\tilde{\eta}_j} \, \sin^2\left(\tilde\vartheta_j-\phi_{1,j}\right)
        \, ,
    \end{align*}
with $\tilde{\mu}_j$, $\tilde{\eta}_j$, and $\tilde\vartheta_j$
satisfying
    \begin{align*}
        \V{F}(\tilde{\mu}_j, \tilde{\eta}_j, \tilde\vartheta_j) =
        \JA{\V{p}_j} + \!\!\!\! \sum_{k\in\NA\backslash\{1,j\}}2\,\C_{jk}
        \, .
    \end{align*}
\end{IEEEproof}

\section{Proof of the Scaling Laws}\label{apd:scaling_laws}


\begin{lem}\label{lem:outage_prob}
Let $\phi_i$'s be $N$ i.i.d. random variables with uniform distribution
over $[\, 0, \, 2\pi)$. Then, for any $0 < \epsilon \leq 1$, there
exist an ${N}_0 \in \mathbb{N}$, such that $\forall \, N > {N}_0$,
    \begin{align}\label{eq:lemma_statement}
        \mathbb{P}\left\{ \sum_{k=1}^N \sum_{j=1}^N
        \sin^2(\phi_k - \phi_j) < \frac{N^2}{32}\right\} < \epsilon
        \, .
    \end{align}
\end{lem}

\begin{IEEEproof}
First, we note that replacing $\phi_i$ with $\phi_i\!\! \mod \pi$
preserves the value of $\sin^2(\phi_k - \phi_j)$. Hence, we can
consider $\phi_i$'s to be i.i.d. and uniformly distributed in $[\,0,
\, \pi )$.

We order the $N$
$\phi_i$'s, such that $0 \leq \phi_{(1)} \leq \phi_{(2)} \leq \cdots \leq \phi_{(\!N\!)} < \pi$. Using order statistics \cite{ArnBlaNag:92}, we find that the joint PDF of the $\phi_{(i)}$'s is
    \begin{align}\label{eq:apd_orderstat_all}
        f(\phi_{(1)},\phi_{(2)}, \ldots, \phi_{(\!N\!)}) = \frac{N!}{\pi^N}
        \; \mathds{1}_{\{0 \leq \phi_{(1)} \leq \phi_{(2)} \leq \cdots \leq \phi_{(\!N\!)} <\pi\}}
        \, ,
    \end{align}
where $\mathds{1}$ is the indicator function. From \eqref{eq:apd_orderstat_all}, the marginal PDF of $\phi_{(k)}$ can be derived as \cite{ArnBlaNag:92}
    \begin{align*}
        f_{\phi_{(\!k\!)}}(x) = \!
        \frac{1}{\pi^N} \frac{N!}{(k-1)! \, (N-k)!} \: x^{k-1} ({\pi} -x)^{N-k}  \mathds{1}_{\{0 \leq x < {\pi}\}}
        .
    \end{align*}
Now consider a large $N=8K$ for some integer $K$, and let $\delta
\triangleq \pi/6$. The function $f_{\phi_{(\!K\!)}}(x)$ has a
maximum at $x=\pi/8$, and is monotonically decreasing in $[\, \pi/8,
\, \pi ) \supset [\, \delta, \, \pi )$. Therefore, we have
    \begin{align}\label{eq:Apd_Prob_phi}
        \mathbb{P} \left\{\phi_{(\!K\!)} > \delta \right\}
        \leq (\pi - \delta) \, f_{\phi_{(\!K\!)}}(\delta)
        \, .
    \end{align}
Since $\lim_{K\rightarrow \infty} f_{\phi_{(\!K\!)}}(\delta) = 0$,
there exists $K_1 \in \mathbb{N}$ such that $\mathbb{P}
\left\{\phi_{(\!K\!)} > \delta \right\} < \epsilon/4$, $\forall K>
K_1$. Note also that
    \begin{align*}
        \mathbb{P} \left\{\phi_{(7K+1)} < \pi - \delta \right\}
        \leq (\pi - \delta) \, f_{\phi_{(7K+1)}}(\pi-\delta)
        \, ,
    \end{align*}
and hence, for the same $K_1$, $\mathbb{P} \left\{\phi_{(7K+1)} <
\pi - \delta \right\} < \epsilon/4$, $\forall K> K_1$. Similar
arguments show that there exists $K_2 \in \mathbb{N}$ such that
$\mathbb{P} \left\{\phi_{(3K+1)} < \pi/2 - \delta \right\} <
\epsilon/4$ and $\mathbb{P} \left\{\phi_{(5K)} > \pi/2 + \delta
\right\} < \epsilon/4$, $\forall K> K_2$.

Combining the above results, we have with a probability
$1-\epsilon$,
    \begin{align*}
        \phi_{(j)} \in \begin{cases}
                \left[\,0, \, \delta \,\right], &   j=1,\ldots,K\,, \\
                \left[\,\pi/2 - \delta , \, \pi/2 + \delta\, \right], &  j = 3K+1,\ldots,5K\,,\\
                \left[\,\pi-\delta , \, \pi \right)\,, & j=7K+1,\ldots, N\,,
            \end{cases}
    \end{align*}
when $K > \max\left\{K_1, K_2 \right\}$.
Therefore,
    \begin{align}\label{eq:lemma_final}
        & \hspace{-4mm} \sum_{k=1}^N \sum_{j=k+1}^N  \sin^2(\phi_{(k)} - \phi_{(j)}) \nonumber \\
        & \geq \sum_{k=1}^K \sum_{j=3K+1}^{5K}
        \sin^2(\phi_{(k)} - \phi_{(j)}) \nonumber \\
        & \hspace{4mm} + \sum_{k=3K+1}^{5K} \sum_{j=7K+1}^{8K} \sin^2(\phi_{(k)} - \phi_{(j)})  \nonumber \\
        & \stackrel{p}{\geq} \left( \sum_{k=1}^K \sum_{j=3K+1}^{5K} \! 1
        + \sum_{k=3K+1}^{5K} \sum_{j=7K+1}^{8K} \! 1\right) \sin^2
        (\frac{\pi}{2} - 2\delta) \nonumber \\
        & = K^2 ,
    \end{align}
where $\stackrel{p}{\geq}$ denotes an inequality with probability approaching one as $K \rightarrow \infty$. Substituting $N=8K$, and noting that the summation in \eqref{eq:lemma_final} considers only half the terms (with $j>k$), we arrive at \eqref{eq:lemma_statement}.

Moreover, the probability in \eqref{eq:Apd_Prob_phi} decreases exponentially with $K$, because if letting $a_K \triangleq f_{\phi_{(\!K\!)}}(\delta)$,
    \begin{align}\label{eq:Apd_outprob_phi1}
        \lim_{K\rightarrow\infty} \frac{a_{K+1}}{a_K}
        & = \frac{(8K+8)(8K+7)\cdots(8K+1)}{(7K+7)(7K+6)\cdots(7K+1) K}
        \,\frac{1}{6} \,\left(\frac{5}{6}\right)^7 \nonumber \\
		& < 1
        \,,
    \end{align}
and hence one can see that $\epsilon$ in \eqref{eq:lemma_statement} decreases exponentially with $N$.
\end{IEEEproof}

\begin{lem}\label{lem:outage_prob2}
Let $\lambda_i$'s be $N$ i.i.d. random variables with arbitrary distribution on the support $[\, 0, \, \lambda_\text{max} \,]$. If $\mathbb{P} \left\{\lambda_i \leq \lambda_0\right\} \leq \epsilon < 1/2$ for some $\lambda_0 \in [\, 0, \, \lambda_\text{max} \,]$, then
\begin{align}\label{eq:lemma_statement2}
    \mathbb{P}\left\{ \lambda_{(N/2+1)} \leq  \lambda_0 \right\} < \tilde\epsilon^N
    \, ,
\end{align}
where $\lambda_{(i)}$ is the order statistics of $\lambda_i$ such that $0 \leq \lambda_{(1)} \leq \lambda_{(2)} \leq \cdots \leq \lambda_{(\!N\!)}$, and $\tilde\epsilon = \sqrt{4\, \epsilon\, (1- \epsilon)}$.
\end{lem}

\begin{IEEEproof}
Denote the probability density and distribution of $\lambda_i$ by $f_\lambda$ and $F_\lambda$, respectively. Consider $N=2K$ for some integer $K$ and $x\in [\, 0, \, \lambda_\text{max} \,]$ such that  $F_\lambda(x)< 1/2$. Using the order statistics, we have
\begin{align*}
	F_{\lambda_{(K+1)}}(x) 
	& = \sum_{j=K+1}^{N} \binom{N}{j} F_\lambda(x)^{j} \left(1-F_\lambda(x)\right)^{N-j} \\
	& \leq \sum_{j=K+1}^{N} 2^N \, F_\lambda(x)^{j} \left(1-F_\lambda(x)\right)^{N-j} \\
	& < 2^N\left(1-F_\lambda(x)\right)^N \sum_{j=K+1}^{\infty} \left( \frac{F_\lambda(x)}{1-F_\lambda(x)} \right)^j \\
	& = \frac{F_\lambda(x)}{1-F_\lambda(x)} {\Big[} 4 \,F_\lambda(x) \left(1 - F_\lambda(x)\right) {\Big]}^K \\
	& < {\Big(} \sqrt{4 \,F_\lambda(x) \left(1 - F_\lambda(x)\right)} {\Big)}^N\,,
\end{align*}
where the first inequality follows from $\binom{N}{j}\leq 2^N$, the second inequality is due to the extension of finite summation, and the last inequality follows from $F_\lambda(x)< 1/2$. Replacing $x$ with $\lambda_0$ gives (\ref{eq:lemma_statement2}).
\end{IEEEproof}

\subsection{Proof of Theorem \ref{thm:Scaling_Law1}}
\label{sec:Apd_scaling_thm}

\begin{IEEEproof} 
We consider first the non-cooperative case, followed by the cooperative case. In either case, without loss of generality, we focus on the first agent at position $\V{p}_1$.

\emph{Non-cooperative case:}
We will show that $\PEB(\V{p}_1) \in {\Omega}({1}/{\Nb})$ and
$\PEB(\V{p}_1) \in O({1}/{\Nb})$, which implies that $\PEB(\V{p}_1)
\in {\Theta}({1}/{\Nb})$.\footnote{Similar to the definition of notation $\Theta(f(n))$, the notation $g(n) \in {\Omega}(f(n))$ and $g(n) \in O(f(n))$ denote, respectively, that $g(n)$ is bounded below by $c_1 f(n)$ and above by $c_2 f(n)$ with probability approaching one as $n \rightarrow \infty$, for some constant $c_1$ and $c_2$.}

For an amplitude loss exponent $b$, signal powers decay with the distance following $\mathsf{SNR}(r)\propto 1/r^{2b}$. We can express the RII from a node at distance $r$ as
    \begin{align*}
        \lambda(r) = \frac{Z}{r^{2b}} \, \mathds{1}_{\{ r_0 \leq r\leq r_\text{max}  \}}
        \,,
    \end{align*}
where $r_0$ is the minimum distance between nodes determined by the
node's physical size, $r_\text{max}$ is the maximum distance between nodes determined by the fixed area associated with dense network setting, and random variable $Z$ accounts for the large- and small-scale fading. Since $0\leq Z \leq z_1$ for some $z_1 \in \mathbb{R}^+$, there exists $z_0 \in (0,z_1)$ such that $\mathbb{P}\left\{ Z \leq z_0\right\} \leq \epsilon_z$ for a given $\epsilon_z\in(0,1)$. Thus, the RII from the $j$th anchor is bounded as $0 <\lambda_\text{min} \leq \lambda_{1,j} \leq \lambda_\text{max}$ with probability 
\begin{align*}
	\mathbb{P}\left\{\lambda_\text{min} \leq \lambda_{1,j} \leq \lambda_\text{max}\right\} \leq 1 - \epsilon_z \, ,
\end{align*}
where $\lambda_\text{min} = z_0/r_\text{max}^{2b}$ and $\lambda_\text{max} = z_1/r_0^{2b}$.

On one hand, we have 
\begin{align}\label{eq:lambda_inequa1}
    \JE{\V{p}_1} \preceq \lambda_\text{max} \sum_{j \in \NB} \R(\phi_{1,j})
    \, ,
\end{align}
By the Cauchy-Schwarz inequality, we have $$\T\left\{[\JE{\V{p}_1}]^{-1}\right\}\cdot \T\left\{\JE{\V{p}_1}\right\} \geq 4 \,.$$ Since the inequality (\ref{eq:lambda_inequa1}) together with the fact that $\T\{{\R(\phi_{1,j})}\}=1$ imply that $\T\left\{\JE{\V{p}_1}\right\} \leq \lambda_\text{max}\, \Nb$, we have
that
\begin{align*}
	\PEB(\V{p}_1) = \T\left\{ \left[\, \JE{\V{p}_1}\,\right]
\right\}^{-1} \geq 4/(\lambda_\text{max} \Nb) \,.
\end{align*}
Therefore, $\PEB(\V{p}_1) \in {\Omega}({1}/{\Nb})$.

On the other hand, for the lower bound, we first order the $\Nb$ RII $\lambda_{1,j}$'s, and then the probability of $\lambda_{(\Nb/2+1)} \leq \lambda_\text{min}$ is exponentially small by Lemma \ref{lem:outage_prob2}, i.e.,
\begin{align}\label{eq:outage_prob2}
	\mathbb{P}\left\{ \lambda_{(\Nb/2+1)} \leq \lambda_\text{min}\right\} \leq \tilde\epsilon^{\Nb} \,,
\end{align}
for some constant $\tilde\epsilon \in (0,1)$. Let $\mathcal{N}_\text{b}'$ denote the set of anchors with RII $\lambda_{(j)}$ such that $j\geq \Nb/2+1$, and we have that
\begin{align}\label{eq:lambda_inequa}
    \mathbb{P}\left\{ \lambda_\text{min} \sum_{j \in \mathcal{N}_\text{b}'} \R(\phi_{1,j}) \preceq \JE{\V{p}_1} \right\} \geq 1 - \epsilon_1 \, ,
\end{align}
where the outage probability $\epsilon_1$ decreases exponentially with $\Nb$. Moreover, since
	\begin{align}\label{eq:Apd_speb_trace}
        \T \left\{  \left[ \sum_{j \in \mathcal{N}_\text{b}'} \R(\phi_{1,j}) \right]^{-1}
        \right\} = \frac{2 \Nb/2}{\sum_{k \in \mathcal{N}_\text{b}'} \sum_{j \in \mathcal{N}_\text{b}'}
        \sin^2(\phi_{1,k}-\phi_{1,j})}
        \, ,
    \end{align}
applying Lemma \ref{lem:outage_prob} gives
    \begin{align}\label{eq:outage_prob}
        \mathbb{P}\left(  \frac{1}{\lambda_\text{min}} \, \T \left\{
        \left[ \sum_{j \in \mathcal{N}_\text{b}'} \R(\phi_{1,j}) \right]^{-1} \right\}
        \leq \frac{128}{\lambda_\text{min} \, \Nb} \right) \geq 1- \epsilon_2
        \, ,
    \end{align}
for sufficiently large $\Nb$. The inequality in (\ref{eq:lambda_inequa}) implies that $$\PEB(\V{p}_1) \leq \frac{1}{\lambda_\text{min}}\T \left\{
\left[ \sum_{j \in \mathcal{N}_\text{b}'} \R(\phi_{1,j}) \right]^{-1} \right\} \,, $$
and hence $\PEB(\V{p}_1) \leq 128/(\lambda_\text{min}\,\Nb)$ with probability approaching one as $\Nb \rightarrow \infty$. Therefore, $\PEB(\V{p}_1) \in O({1}/{\Nb})$ with probability 1. 

Note that since both the outage probability $\epsilon_1$ in (\ref{eq:lambda_inequa}) and $\epsilon_2$ in  (\ref{eq:outage_prob}) decrease exponentially with $\Nb$, the outage probability $\epsilon(\Nb)$ of the scaling law in (\ref{eq:def_scaling}) decreases exponentially with $\Nb$.



\emph{Cooperative case:} 
For the cooperative case, we will use the lower and upper approximations of the EFIM from \eqref{eq:Anal_Lower} and \eqref{eq:Anal_Upper}. The upper approximation gives
    \begin{align*}
        \JU{\V{p}_1} & = \JA{\V{p}_1}
        + \!\!\sum_{j\in\NA\backslash\{1\}} \xi_{1,j}^\text{U} \, \C_{1,j} \\
        & \preceq \!\!\sum_{j \in \NB \cup \NA\backslash\{1\}} \lambda_{1,j}
        \, \R(\phi_{1,j}) \, ,
    \end{align*}
where the inequality is obtained by treating all other agents to be
anchors, i.e., $\xi_{1,j}^\text{U}=1\;(j\in\NA)$. In this case, there
are equivalently $\Nb+\Na-1$ anchors, and similar analysis as in the
non-cooperative case shows that $\PEB(\V{p}_1) \in
\Omega({1}/{(\Nb+\Na)})$.

On the other hand, from the lower approximation, we have, with probability approaching one, that
    \begin{align}\label{eq:JL_SCALING}
        \JL{\V{p}_1} & = \JA{\V{p}_1}
        + \!\!\sum_{j\in\NA\backslash\{1\}} \xi_{1,j}^\text{L} \, \C_{1,j} \nonumber \\
        & \succeq  \; \tilde{\nu} \!\!\sum_{j \in \NB \cup \NA\backslash\{1\}} \R(\phi_{1,j})
        \, ,
    \end{align}
where $\tilde{\nu}>0$ is a given lower bound on both the RII  $\lambda_{1,j}\;(j \in \NB)$ and the effective RII $\xi_{1,j}^\text{L}\,\nu_{1,j}\;(j\in\NA)$. From Lemma \ref{lem:outage_prob2}, we can find such $\tilde{\nu}$ for the dense network setting, because there exist constants $0<c_1,c_2<+\infty$ such that $\lambda_{1,j}>c_1$, $\nu_{1,j}>c_1$, and $\Delta_j(\phi_{1,j})< c_2$ with probability approaching one; defining $\tilde{\nu} \triangleq c_1/(1+c_1\cdot c_2)$ implies $\lambda_{1,j}\geq  \tilde{\nu}$ and $\xi_{1,j}^\text{L}\,\nu_{1,j} \geq \tilde{\nu}$ since $\xi_{1,j}^\text{L} = \left[\, 1+ \Delta_j(\phi_{1,j})\nu_{1,j}\, \right]^{-1}$.
Applying Lemma \ref{lem:outage_prob} and \ref{lem:outage_prob2}, and following a similar line of reasoning as in the non-cooperative case, we find $\PEB(\V{p}_1) \in O(1/(\Nb+\Na))$ with probability approaching one as $\Nb,\Na \rightarrow +\infty$. Thus, we conclude that the SPEB in cooperative networks scales as $\Theta({1}/{(\Nb+\Na)})$.
\end{IEEEproof}

\subsection{Proof of Theorem \ref{thm:Scaling_Law2}}
\label{sec:Apd_scaling_thm2}

\begin{IEEEproof}
Let $\rho_\text{b}$ denote the density of anchor nodes uniformly distributed in an extended network. Consider an area within distance $R$ to agent 1, then the expected number of anchors within that area is $\Nb = \rho_\text{b}\, \pi R^2$. Following a similar analysis leading to (\ref{eq:outage_prob2}), we can show that the effect of large- and small-scale fading together with path-loss on the RII can be bounded as ${c_1}/{r^{2b}} \leq \lambda(r) \leq {c_2}/{r^{2b}}$ for some constants $0<c_1<c_2<+\infty$, with an outage probability exponentially decreasing with $\Nb$ and $\Na$. This implies that, with probability approaching one, the large- and small-scale fading will not affect the scaling law,\footnote{It will be shown that the overall outage is dominated by the spatial topology for a large number of nodes, and thus we can ignore the outage due to fading.} and hence we can consider the RII from a node at distance $r$ as
    \begin{align*}
        \lambda(r) = \frac{1}{r^{2b}} \, \mathds{1}_{\{ r\geq r_0  \}}
        \,,
    \end{align*}
for the analysis of the scaling laws. Since each anchor is uniformly distributed in
the given area, the PDF of the RII can be written as
    \begin{align*}
        f(\lambda) = \frac{1}{b \left( R^2 - r_0^2 \right)}
        \, \lambda^{-\frac{b+1}{b}}
        \, \mathds{1}_{\{{1}/{R^{2b}} \leq \lambda \leq {1}/{r_0^{2b}}\}}
        \, ,
    \end{align*}
with mean
    \begin{align}
        \E \{ \lambda \} 
        = \! \begin{cases}
                \frac{1}{b-1} \frac{r_0^{2-2b} - R^{2-2b}}{R^2 - r_0^2}\,,
                & b > 1\,,\\
                \frac{\ln R^2 - \ln r_0^2}{R^2 - r_0^2}\,,
                & b = 1\,, \\
                \frac{1}{1-b} \frac{R^{2-2b} - r_0^{2-2b}}{R^2 - r_0^2}\,,
                & 0< b < 1 ,
            \end{cases}
    \end{align}
and second moment 
	\begin{align}
	    \E \left\{ \lambda^2 \right\}
	    = \begin{cases}
	            \frac{1}{2b-1} \frac{r_0^{2-4b} - R^{2-4b}}{R^2 - r_0^2} \,,
	            & b > 1\,, \\
	            \frac{1}{r_0^2 \, R^2}\,,
	            & b = 1 \,, \\
	            \frac{1}{2b-1} \frac{r_0^{2-4b} - R^{2-4b}}{R^2 - r_0^2} \,,
	            & 0< b < 1\,.
	        \end{cases}
	\end{align}
Note that $\Nb \propto R^2$, we can show that the mean scales as
\begin{align}\label{eq:lambda_mean}
    \E \{ \lambda \} 
    \in \begin{cases}
            \Theta \left( {1}/{\Nb} \right),
            & b > 1\,,\\
            \Theta \left({\log \Nb}\,/\,{\Nb}\right),
            & b = 1\,, \\
            \Theta \left( 1/N_\text{b}^{b} \right),
            & 0< b < 1 ,
        \end{cases}
\end{align}
and the variance always scales as
\begin{align}\label{eq:lambda_var}
	\mathbb{V}\text{ar} \left\{ \lambda \right\}  \in \Theta \left({1}/{\Nb}\right) \,.
\end{align}

When $b>1$, it follows that, for fixed densities of anchors and
agents, $\T\left\{ \JE{\V{p}_1} \right\} \in \Theta(1)$ with
probability approaching one as $\Nb \rightarrow +\infty$, which
implies that $\PEB(\V{p}_1) \in \Theta(1)$.

We will show that when $b=1$, the $\PEB(\V{p}_1)$ scales as
$\Theta({1}/{\log \Nb})$ and $\Theta({1}/{\log (\Nb+\Na)})$ for
the non-cooperative case and cooperative case, respectively. Using a similar argument, we can easily show that for $0<b<1$ the SPEB scales as $\Theta({1}/N_\text{b}^{b-1})$ and $\Theta({1}/(\Nb+\Na)^{b-1})$ for the non-cooperative case and cooperative case, respectively.

\subsubsection*{Non-cooperative case ($b=1$)}

We introduce a random variable $Y_{\Nb} = \sum_{j \in \NB} \lambda_{1,j} / \log(\Nb)$. From \eqref{eq:lambda_mean} and \eqref{eq:lambda_var},
we have
    \begin{align*}
        \lim_{\Nb \rightarrow \infty}\E \left\{ Y_{\Nb} \right\} = C
        \, ,
    \end{align*}
for some constant $C$, and
    \begin{align*}
        & \hspace{-4mm} \lim_{\Nb \rightarrow \infty}\E \left\{ | Y_{\Nb}-C |^2 \right\} \\
        & = \lim_{\Nb \rightarrow \infty}\mathbb{V}\text{ar} \left\{Y_{\Nb}\right\} 
		+ \lim_{\Nb \rightarrow \infty} | \E \left\{ Y_{\Nb} \right\} -C |^2 \nonumber \\
        & \hspace{5mm} +  \lim_{\Nb \rightarrow \infty} 2\left( \E \left\{ Y_{\Nb} \right\} -C \right)\cdot \E \left\{Y_{\Nb}-\E\left\{Y_{\Nb}\right\}\right\}
        \nonumber \\
        &= 0
        \, .
    \end{align*}
This implies that $\sum_{j \in \NB} \lambda_{1,j}$ scales as $\Theta(\log \Nb)$ with probability approaching one, and hence $\T\left\{ \JE{\V{p}_1} \right\} \in \Theta \left(\log \Nb  \right)$. Using a similar analysis as in Appendix \ref{sec:Apd_scaling_thm}, we can show that $\PEB(\V{p}_1) \in \Omega({1}/{\log \Nb})$.

For the upper bound, using the same argument as in Lemma \ref{lem:outage_prob}, we can show that with probability approaching one, there are $\Nb/8$ anchors with angle $\phi_k \in [\,0,\pi/6\,]$ and $\Nb/8$ anchors with angle $\phi_k \in [\,\pi/3, \pi/2\,]$ to the agent. We denote these two disjoint sets of anchors by $\mathcal{N}_1$ and $\mathcal{N}_2$, and define
    \begin{align*}
        \tilde{\V{J}}_\text{e}(\V{p}_1) & \triangleq
        \sum_{j \in \mathcal{N}_1\cup\mathcal{N}_2} \lambda_{1,j} \, \R(\phi_{1,j})
        \, ,
    \end{align*}
and
    \begin{align*}
        \tilde{\V{J}}_\text{e}^*(\V{p}_1) \triangleq
        \left(\sum_{j \in \mathcal{N}_1} \lambda_{1,j}\right) \R(\pi/6) + \left( \sum_{j \in \mathcal{N}_2} \lambda_{1,j} \right) \R(\pi/3)
        \, .
    \end{align*}
Then, we have
    \begin{align}\label{eq:Apd_extend_1}
        \T \left\{ \left[\JE{\V{p}_1}\right]^{-1} \right\}
        \leq  \T \left\{ \left[\tilde{\V{J}}_\text{e}(\V{p}_1)\right]^{-1} \right\}
        \leq \T \left\{ \left[\tilde{\V{J}}_\text{e}^*(\V{p}_1)\right]^{-1} \right\}  ,
    \end{align}
where the first inequality comes from $\mathcal{N}_1\cup\mathcal{N}_2 \subseteq \NB$, and the second inequality is due to the fact that the SPEB increases if we set $\phi_{1,j}=\pi/6$ for $j\in\mathcal{N}_1$ and $\phi_{1,j}=\pi/3$ for $j\in\mathcal{N}_2$.\footnote{This can be seen from \eqref{eq:Apd_speb_trace} that every element in the sum of the denominator decreases if letting $\phi_{1,j}=\pi/6$ for $j\in\mathcal{N}_1$ and $\phi_{1,j}=\pi/3$ for $j\in\mathcal{N}_2$.} Since both $\sum_{j \in \mathcal{N}_1}\lambda_{1,j}$ and $\sum_{j \in \mathcal{N}_2} \lambda_{1,j}$ scale as $\Theta(\log\Nb)$, $\PEB(\V{p}) \in O({1}/{\log \Nb})$ with probability approaching one. Therefore, the SPEB in non-cooperative extended networks scales as $\Theta({1}/{\log \Nb})$.

We finally check the probability of outage, i.e., $\sum_{j\in\NB}\lambda_{1,j}$ is not in $\Theta(\log \Nb)$. For a fixed large $\Nb$, the distribution of $\sum_{j\in\NB}\lambda_{1,j}/\sqrt{\Nb}$ can be approximated as the normal distribution $N(\log \Nb/\sqrt{\Nb}, 1/\Nb)$, and hence\footnote{The notation $\cong$ denotes ``on the order of.''}
    \begin{align}\label{eq:Apd_outprob_lambda1}
       \mathbb{P} {\bigg(} {\Big|} \sum_{j\in\NB}\lambda_{1,j} & - \log \Nb {\Big|} > \frac{1}{2} \log \Nb {\bigg)} \nonumber \\
        & = 2 \,Q\left( \frac{1}{2} \log\Nb\right) \nonumber \\
        & \cong \frac{1}{\log\Nb} \exp \left\{-\frac{1}{8}\log^2\Nb\right\}
        \, ,
    \end{align}
where $Q(\cdot)$ is the tail probability function of standard normal distribution. Approximations and bounds for the tail probability function can be found in \cite{ConWinChi:J03, GriSti:01,ChiDarSim:03}. Moreover, when $0<b<1$ a similar argument leads to
    \begin{align}\label{eq:Apd_outprob_lambda2}
        \mathbb{P}  {\bigg(} {\Big|} \sum_{k\in\NB}\lambda_k & - N_\text{b}^{1-b} {\Big|} >  \frac{1}{2}\,N_\text{b}^{1-b}  {\bigg)} \nonumber \\
        & = 2\,Q \left(\frac{1}{2}\,N_\text{b}^{1-b} \right) \nonumber \\
        & \cong \frac{1}{N_\text{b}^{1-b}} \exp \left\{-\frac{1}{8} N_\text{b}^{2-2b}\right\}
        \, .
    \end{align}


\subsubsection*{Cooperative case ($b=1$)}

The cooperative case can be proved similar to the above non-cooperative case in conjunction with the cooperative case of Theorem \ref{thm:Scaling_Law1}. It turns out that the SPEB can be shown to scale as $\Omega({1}/{\log (\Nb+\Na)})$ when all other agents are considered to be anchors. We can also show that, with probability approaching one, the SPEB scales as $O({1}/{\log (\Nb+\Na)})$, using the lower approximation of the EFIM, and an argument similar to \eqref{eq:Apd_extend_1}.
%
%
\end{IEEEproof}



\bibliographystyle{IEEEtran}

\begin{IEEEbiographynophoto}
{Yuan Shen} (S'05) received his B.S.\ degree (with highest honor) from Tsinghua University, China, in 2005, and S.M.\ degree from the Massachusetts Institute of Technology (MIT) in 2008, both in electrical engineering. 

Since 2005, he has been with Wireless Communications and Network Science Laboratory at MIT, where he is now a Ph.D.\ candidate. He was with the Hewlett-Packard Labs, CA, in winter 2009, the Corporate R\&D of Qualcomm Inc., CA, in summer 2008, and the Intelligent Transportation Information System Laboratory, Tsinghua University, China, from 2003 to 2005. His research interests include communication theory, information theory, and statistical signal processing. His current research focuses on wideband localization, cooperative networks, and ultra-wide bandwidth communications.

Mr.~Shen served as a member of the Technical Program Committee (TPC) for the IEEE Global Communications Conference (GLOBECOM) in 2010, 
the IEEE International Conference on Communications (ICC) in 2010, and the IEEE Wireless Communications \& Networking Conference (WCNC) in 2009 and 2010. He received the Ernst A.~Guillemin Thesis Award (first place) for the best S.M.~thesis from the Department of Electrical Engineering and Computer Science at MIT in 2008, the Roberto Padovani Scholarship from Qualcomm Inc.~in 2008, the Best Paper Award from the IEEE WCNC in 2007, and the Walter A.~Rosenblith Presidential Fellowship from MIT in 2005.
\end{IEEEbiographynophoto}

\begin{IEEEbiographynophoto}
{Henk Wymeersch} (S'98-M'05) is Assistant Professor with the
Department of Signals and Systems at Chalmers University, Sweden.
Prior to joining Chalmers, he was a postdoctoral associate with the
Laboratory for Information and Decision Systems (LIDS) at the
Massachusetts Institute of Technology (MIT). Henk Wymeersch obtained
the Ph.D. degree in Electrical Engineering / Applied Sciences in 2005
from Ghent University, Belgium. He is author of the book "Iterative
Receiver Design" (Cambridge University Press, August 2007). His
research interests include algorithm design for wireless transmission, statistical inference and iterative processing.
\end{IEEEbiographynophoto}


\begin{IEEEbiographynophoto}
{Moe Z. Win} 
(S'85-M'87-SM'97-F'04)
received both the Ph.D.\ in Electrical Engineering and M.S.\ in Applied Mathematics
as a Presidential Fellow at the University of Southern California (USC) in 1998.
He received an M.S.\ in Electrical Engineering from USC in 1989, and a B.S.\ ({\em magna cum laude})
in Electrical Engineering from Texas A\&M University in 1987.

Dr.\ Win is an Associate Professor at the Massachusetts Institute of Technology (MIT).
Prior to joining MIT, he was at AT\&T Research
Laboratories for five years and at the Jet Propulsion Laboratory for seven years.
His research encompasses developing fundamental theories, designing algorithms, and
conducting experimentation for a broad range of real-world problems.
His current research topics include location-aware networks,
time-varying channels, multiple antenna systems, ultra-wide bandwidth
systems, optical transmission systems, and space communications systems.

Professor Win is an IEEE Distinguished Lecturer and
        elected Fellow of the IEEE, cited for ``contributions to wideband wireless transmission.''
He was honored with
        the IEEE Eric E. Sumner Award (2006), an IEEE Technical Field Award for
        ``pioneering contributions to ultra-wide band communications science and technology.''
Together with students and colleagues, his papers have received several awards including
        the IEEE Communications Society's Guglielmo Marconi Best Paper Award (2008)
    and the IEEE Antennas and Propagation Society's Sergei A. Schelkunoff Transactions Prize Paper Award (2003).
His other recognitions include
        the Laurea Honoris Causa from the University of Ferrara, Italy (2008),
        the Technical Recognition Award of the IEEE ComSoc Radio Communications Committee (2008),
        Wireless Educator of the Year Award (2007),
        the Fulbright Foundation Senior Scholar Lecturing and Research Fellowship (2004),
        the U.S. Presidential Early Career Award for Scientists and Engineers (2004),
        the AIAA Young Aerospace Engineer of the Year (2004),
    and the Office of Naval Research Young Investigator Award (2003).

Professor Win has been actively involved in organizing and chairing
a number of international conferences. He served as
    the Technical Program Chair for
        the IEEE Wireless Communications and Networking Conference in 2009,
        the IEEE Conference on Ultra Wideband in 2006,
        the IEEE Communication Theory Symposia of ICC-2004 and Globecom-2000,
        and
        the IEEE Conference on Ultra Wideband Systems and Technologies in 2002;
    Technical Program Vice-Chair for
        the IEEE International Conference on Communications in 2002; and
    the Tutorial Chair for
        ICC-2009 and
        the IEEE Semiannual International Vehicular Technology Conference in Fall 2001.
He was
    the chair (2004-2006) and secretary (2002-2004) for
        the Radio Communications Committee of the IEEE Communications Society.
Dr.\ Win is currently
    an Editor for {\scshape IEEE Transactions on Wireless Communications.}
He served as
    Area Editor for {\em Modulation and Signal Design} (2003-2006),
    Editor for {\em Wideband Wireless and Diversity} (2003-2006), and
    Editor for {\em Equalization and Diversity} (1998-2003),
        all for the {\scshape IEEE Transactions on Communications}.
He was Guest-Editor
        for the
        {\scshape Proceedings of the IEEE}
        (Special Issue on UWB Technology \& Emerging Applications) in 2009 and
        {\scshape IEEE Journal on Selected Areas in Communications}
        (Special Issue on Ultra\thinspace-Wideband Radio in Multiaccess
        Wireless Communications) in 2002.
\end{IEEEbiographynophoto}

\end{document}